\documentclass[usenatbib,useams]{mn2e}

\pdfoutput=1

\usepackage{times}
\usepackage{aas_macros}
\usepackage{graphicx}

\def\ffrac#1#2{\frac{#1}{#2}}

\def\kms{\mathrm{\,km\,s^{-1}}}
\def\Msun{\mathrm{M}_\odot}

\def\sign{\rm sign}

\def\seta{\sin \eta}
\def\setasq{\sin^2 \eta}
\def\ceta{\cos \eta}
\def\edenom{\left[ \vr^2 + \vt^2 \setasq \right] ^{\ffrac12}}

\def\tu{t_{\mathrm{u}}}
\def\etau{\eta_{\mathrm{u}}}

\def\Mand{M_{\mathrm{M31}}}
\def\dand{d_{\mathrm{M31}}}

\def\vcor{v_{\mathrm{cor}}}
\def\vlos{v_{\mathrm{los}}}
\def\vr{v_{\mathrm{r}}}
\def\vt{v_{\mathrm{t}}}

\def\norb{N_{\mathrm{orb}}}
\def\npos{N_{\mathrm{pos}}}
\def\ntot{N_{\mathrm{tot}}}

\def\Msat{M_{\mathrm{sat}}}

\def\vsat{v_{\mathrm{sat}}}
\def\evsat{\sigma_{\mathrm{\vsat}}}

\def\ri{r_{\mathrm{i}}}
\def\vi{v_{\mathrm{i}}}

\def\Thsat{\Theta_{\mathrm{sat}}}
\def\Thi{\Theta_{\mathrm{i}}}
\def\Thsys{\Theta_{\mathrm{sys}}}

\def\rmin{r_{\mathrm{min}}}
\def\rp{r_{\mathrm{p}}}
\def\ra{r_{\mathrm{a}}}

\def\mutrue{\mu_{\mathrm{true}}}
\def\sigsys{\sigma_{\mathrm{sys}}}

\usepackage[pdftex, bookmarks=true]{hyperref}

\hypersetup{
	pdftitle={A census of orbital properties of the M31 satellites},
	pdfauthor={Laura~L.~Watkins, N.~Wyn~Evans, Glenn~van~de~Ven}
}

\def\equationautorefname~#1\null{%
  equation~(#1)\null
}

\title[A census of orbital properties of the M31 satellites]{A census of orbital properties of the M31 satellites}

\author[L.L. Watkins et al.]{Laura~L.~Watkins$^{1,2}$\thanks{watkins@mpia.de},
    N.~Wyn~Evans$^2$, Glenn~van~de~Ven$^1$ \\
    $^1$Max Planck Institute for Astronomy, K\"{o}nigstuhl 17, Heidelberg 69117, Germany \\
    $^2$Institute of Astronomy, University of Cambridge, Madingley Road, Cambridge, CB3 0HA, UK}

\date{Accepted ???; Received ???; Submitted ???}

\pagerange{\pageref{firstpage}--\pageref{lastpage}} \pubyear{2012}

\begin{document}

\label{firstpage}

\maketitle

\begin{abstract}
    We present an analysis of the dynamics of the M31 satellite system.  Proper motion data exist for only two of the M31 satellites. We account for this incompleteness in velocity data by a statistical analysis using a combination of the timing argument and phase-space distribution functions.  The bulk of the M31 satellites are well fit by these models and we offer a table of orbital properties, including period, eccentricity and semi-major axis. This enables us to search for evidence of group infall based on orbital similarity rather than propinquity on the sky. Our results favour an association between Cass\,II and NGC\,185, as the orbital parameters are in close agreement, but not for NGC\,185 and NGC\,147, which have often been associated in the past. Other possible satellite groupings include the pair And\,I and And\,XVII; the pair And\,IX and And\,X; and the triple And\,V, And\,XXV and NGC\,147.  And\,XXII has been claimed as a satellite of M33; we find that they are not moving independently along the same orbit, but cannot determine whether they are orbiting each other or are unrelated.  Two satellites, And\,XII and And\,XIV, have high line-of-sight velocities, consistent with very recent infall from the edge of the Local Group. They are not well described by our underlying smooth phase space distribution function, and are reanalysed without priors on their orbital parameters.  For And\,XIV, multiple pericentric passages are possible and improved distance information is needed to draw further conclusions. For And\,XII, orbits which assume at least one pericentric passage can be ruled out and it must be on its first infall into the M31 system.
\end{abstract}



\section{Introduction}
\label{sect:introduction}

In this paper, we present an analysis of the distances and velocities of the satellites of M31 within the framework of the timing argument.  There have been a number of wide field surveys of M31 in recent years that have more than doubled the number of satellites \citep{zucker2004, zucker2007, martin2006, majewski2007, ibata2007, irwin2008, mcconnachie2008, martin2009, richardson2011, slater2011, bell2011, irwin2012}. Spectroscopic follow-up has provided radial velocities for almost all the new discoveries \citep[e.g.][]{tollerud2012, collins2012} and the existing dataset is now substantial. However, proper motion data are not currently measurable with existing optical telescopes. For only two of the M31 satellites (M33 and IC\,10) can proper motions be derived using VLBI observations of maser emission from star-forming regions.  Although future, deeper surveys will discover many more such galaxies (e.g., SkyMapper, LSST), now seems an opportune moment to analyse the orbital properties of the known satellites.

The M31 galaxy is believed to have had a more violent accretion history than the Milky Way. Its stellar halo is more dense and extended than that of the Milky Way, its collection of satellite galaxies and globular clusters more numerous.  Two of the entourage, And\,XII \citep{martin2006, chapman2007, collins2010} and And\,XIV \citep{majewski2007, kalirai2010}, are unusual in that they have extreme kinematics with high line-of-sight velocities. And\,XII may have been formed outside of the virial radius of the Local Group \citep{chapman2007} and only recently been accreted.  Similarly, \citet{majewski2007} found that, even at its projected radius, And\,XIV is nearly at the M31 escape velocity, which implies that And\,XIV may also be on its first infall into the Local Group. Hierarchical models of formation can often produce dSphs on eccentric orbits at late times, as the smaller galaxies begin to interact with the nascent, central galaxy. For example, \citet{sales2007} suggested that infall of a satellite pair, followed by a three-body encounter may detach the lighter companion and eject it on to an eccentric orbit. Such satellites may be identifiable as outliers in a census of orbital properties.

The main hurdle is that, for most of the M31 satellites, only the line-of-sight velocities are known; there are no proper motions.  This means that statistical techniques have to be developed to associate characteristic orbital periods, eccentricities, semi-major axes and numbers of pericentric passages with each satellite galaxy. We show how to do this with a novel combination of the timing argument and phase-space distribution functions.  The principle behind the timing argument is simple: for two galaxies that are nearby and moving towards each other at the present time, we assume that, at the time of the Big Bang, they were close together and began to move apart with the Hubble expansion.  At high redshift, their mutual gravitational attraction would have caused them to decouple from the Hubble flow and their orbit can then be described using Newtonian dynamics.  Given their present separation and velocities and the time for which they have been interacting, we can estimate the properties of their orbit.

\citet{kahn1959} first applied the timing argument to the relative motion of the Milky Way and M31 to infer that there must be a significant amount of intergalactic mass in the Local Group.  This analysis has been re-visited in the intervening years \citep[e.g.][]{vandermarel2008} and, more recently, it has been applied to certain of the Milky Way satellites in order to estimate the Milky Way mass \citep{li2008a, kallivayalil2009}.  Of course, the timing argument treats the two galaxies as point masses, which is clearly an oversimplification. Galaxies are extended objects, and have themselves evolved and accreted a significant amount of mass since the Big Bang.  \citet{li2008a} studied analogues of the Local Group in the Millennium Simulation to show that the timing argument gives an almost unbiased estimate of the combined mass of two large galaxies, like the Milky Way and M31.  They then extended the analysis to a large galaxy with a smaller satellite and showed that it is equally effective for determining the host mass.  So it seems that, despite the rather generous assumptions, the timing argument is surprisingly effective.

\autoref{sect:data} introduces the data for the M31 satellites. \autoref{sect:mc_sims} introduces a distribution of semi-major axes and eccentricities for a satellite galaxy population in the outer parts of M31, and shows how to fix the parameters via a maximum likelihood method. \autoref{sect:results} implements the algorithm, and provides maximum likelihood solutions for the whole dataset, as well as individual satellites.  For each satellite, we provide the most likely orbital period, eccentricity, semi-major axis in \autoref{table:orbprops}.  In \autoref{sect:infall}, we identify possible candidates for group infall.  \autoref{sect:minmass} relaxes some of the assumptions we make earlier in the paper and reanalyses a number of satellites that are not well fit by the earlier models; seeking minimum mass solutions to the Timing Argument to account for the lack of proper motion data.  Finally, we conclude in \autoref{sect:conclusions}.


\section{Data}
\label{sect:data}

\begin{table*}
    \caption{Data table for the Local Group galaxies used in this analysis.}
    \label{table:data}
    \begin{tabular}{lccccc}
        \hline
        \hline
        & $\alpha_{0,\mathrm{J2000}}$ & $\delta_{0,\mathrm{J2000}}$
            & $d_{\rm h}$ (kpc) & $v_{\rm h}$ (km/s)
            & sources $^a$ \\
        \hline
        M31 & 00 42 44.3 & +41 16 09.0 & $779^{+19}_{-18}$ & $-301.0 \pm 1.0$ & 1,2,3 \\
        And\,I & 00 45 39.8 & +38 02 28.0 & $727^{+18}_{-17}$ & $-376.3 \pm 2.2$ & 1,2,14 \\
        And\,II & 01 16 29.8 & +33 25 09.0 & $630 \pm 15$ & $-193.6 \pm 1.0$ & 1,2,14 \\
        And\,III & 00 35 33.8 & +36 29 52.0 & $723^{+18}_{-24}$ & $-344.3 \pm 1.7$ & 1,2,14 \\
        And\,V & 01 10 17.1 & +47 37 41.0 & $742^{+21}_{-22}$ & $-397.3 \pm 1.5$ & 1,2,15 \\
        And\,VI (Peg dSph) & 23 51 46.3 & +24 34 57.0 & $783 \pm 25$ & $-340.8 \pm 1.9$ & 1,1,14 \\
        And\,VII (Cass\,I) & 23 26 31.0 & +50 41 31.0 & $763 \pm 35$ & $-307.2 \pm 1.3$ & 1,1,14 \\
        And\,IX & 00 52 51.1 & +43 11 48.6 & $600^{+91}_{-23}$ & $-209.4 \pm 2.5$ & 4,2,15 \\
        And\,X & 01 06 35.3 & +44 48 03.8 & $670^{+24}_{-39}$ & $-164.1 \pm 1.7$ & 5,2,14 \\
        And\,XI & 00 46 21.0 & +33 48 22.0 & $763^{+29}_{-106}$ & $-427.0^{+2.9}_{-2.8}$ & 4,2,14 \\
        And\,XII & 00 47 27.0 & +34 22 29.0 & $928^{+40}_{-136}$ & $-557.1 \pm 1.7$ & 4,2,14 \\
        And\,XIII & 00 51 51.0 & +33 00 16.0 & $760^{+126}_{-154}$ & $-185.4 \pm 2.4$ & 4,2,15 \\
        And\,XIV & 00 51 35.0 & +29 41 49.0 & $793^{+23}_{-179}$ & $-480.6 \pm 1.2$ & 4,2,14 \\
        And\,XV & 01 14 18.7 & +38 07 03.0 & $626^{+79}_{-35}$ & $-323.0 \pm 1.4$ & 4,2,15 \\
        And\,XVI & 00 59 29.8 & +32 22 36.0 & $476^{+44}_{-29}$ & $-367.3 \pm 2.8$ & 4,2,15 \\
        And\,XVII & 00 37 08.0 & +44 18 53.0 & $727^{+39}_{-25}$ & $-251.1^{+1.5}_{-1.6}$ & 5,2,14 \\
        And\,XVIII & 00 02 14.5 & +45 05 20.0 & $1214^{+40}_{-43}$ & $-332.1 \pm 2.7$ & 6,2,15 \\
        And\,XIX & 00 19 32.1 & +35 02 37.1 & $821^{+32}_{-148}$ & $-111.2^{+1.2}_{-1.3}$ & 6,2,14 \\
        And\,XX & 00 07 30.7 & +35 07 56.4 & $741^{+42}_{-52}$ & $-456.2^{+3.0}_{-3.4}$ & 6,2,14 \\
        And\,XXI & 23 54 47.7 & +42 28 15.0 & $827^{+23}_{-25}$ & $-362.7 \pm 0.8$ & 7,2,14 \\
        And\,XXII (Tri\,I) & 01 27 40.0 & +28 05 25.0 & $920^{+32}_{-139}$ & $-129.0^{+2.1}_{-2.2}$ & 7,2,15 \\
        And\,XXIII & 01 29 21.8 & +38 43 08.0 & $748^{+31}_{-21}$ & $-242.7 \pm 1.0$ & 8,2,14 \\
        And\,XXIV & 01 18 30.0 & +46 21 58.0 & $898^{+28}_{-42}$ & $-127.8^{+5.3}_{-5.4}$ & 8,2,14 \\
        And\,XXV & 00 30 08.9 & +46 51 07.0 & $736^{+23}_{-69}$ & $-107.8^{+1.0}_{-0.9}$ & 8,2,14 \\
        And\,XXVI & 00 23 45.6 & +47 54 58.0 & $754^{+218}_{-164}$ & $-260.6^{+4.0}_{-3.7}$ & 8,2,14 \\
        And\,XXVII & 00 37 27.1 & +45 23 13.0 & $1255^{+42}_{-474}$ & $-534.8^{+5.4}_{-4.9}$ & 8,2,14 \\
        Cass\,II (And\,XXX) & 00 36 34.9 & +49 38 48.0 & $681^{+32}_{-78}$ & $-141.4^{+5.8}_{-6.7}$ & 9,2,14 \\
        IC\,10 & 00 20 24.5 & +59 17 30.0 & $660 \pm 66$ & $-348.0 \pm 1.0$ & 10,11,16 \\
        IC\,1613 & 01 04 54.1 & +02 07 60.0 & $715 \pm 40$ & $-232.0 \pm 1.0$ & 10,12,16 \\
        M32 & 00 42 42.1 & +40 51 59.0 & $781 \pm 20$ & $-200.0 \pm 6.0$ & 10,13,16 \\
        M33 & 01 33 50.9 & +30 39 36.0 & $820^{+20}_{-19}$& $-180.0 \pm 1.0$ & 1,2,16 \\
        NGC\,147 & 00 33 12.1 & +48 30 32.0 & $712^{+21}_{-19}$ & $-193.0 \pm 3.0$ & 1,2,17 \\
        NGC\,185 & 00 38 58.0 & +48 20 15.0 & $620^{+19}_{-18}$ & $-202.0 \pm 3.0$ & 1,2,17 \\
        NGC\,205 & 00 40 22.1 & +41 41 07.0 & $824 \pm 27$ & $-241.0 \pm 3.0$ & 1,1,17 \\
        Pegasus & 23 28 23.2 & +14 44 35.0 & $919 \pm 30$ & $-184.5 \pm 0.3$ & 1,1,18 \\
        Pisces & 01 03 52.9 & +21 53 05.0 & $769 \pm 23$ & $-286.5 \pm 0.3$ & 1,1,18 \\
        \hline
        \hline
    \end{tabular}
    
    \raggedright
    \medskip
    \textbf{Notes}: $^a$ sources are given in turn for position, distance, line-of-sight velocity \\
    \textbf{Sources}:
        (1) \citet{mcconnachie2005};
        (2) \citet{conn2012};
        (3) \citet{courteau1999};
        (4) \citet{collins2010};
        (5) \citet{brasseur2011};
        (6) \citet{mcconnachie2008};
        (7) \citet{martin2009};
        (8) \citet{richardson2011};
        (9) \citet{irwin2012};
        (10) \citet{karachentsev2004};
        (11) \citet{sakai1999};
        (12) \citet{cole1999};
        (13) see \autoref{sect:distm32};
        (14) \citet{collins2012};
        (15) \citet{tollerud2012};
        (16) \citet{huchra1999};
        (17) \citet{bender1991};
        (18) \citet{huchtmeier2003}
\end{table*}

\begin{figure}
\begin{center}
    \includegraphics[width=\linewidth]{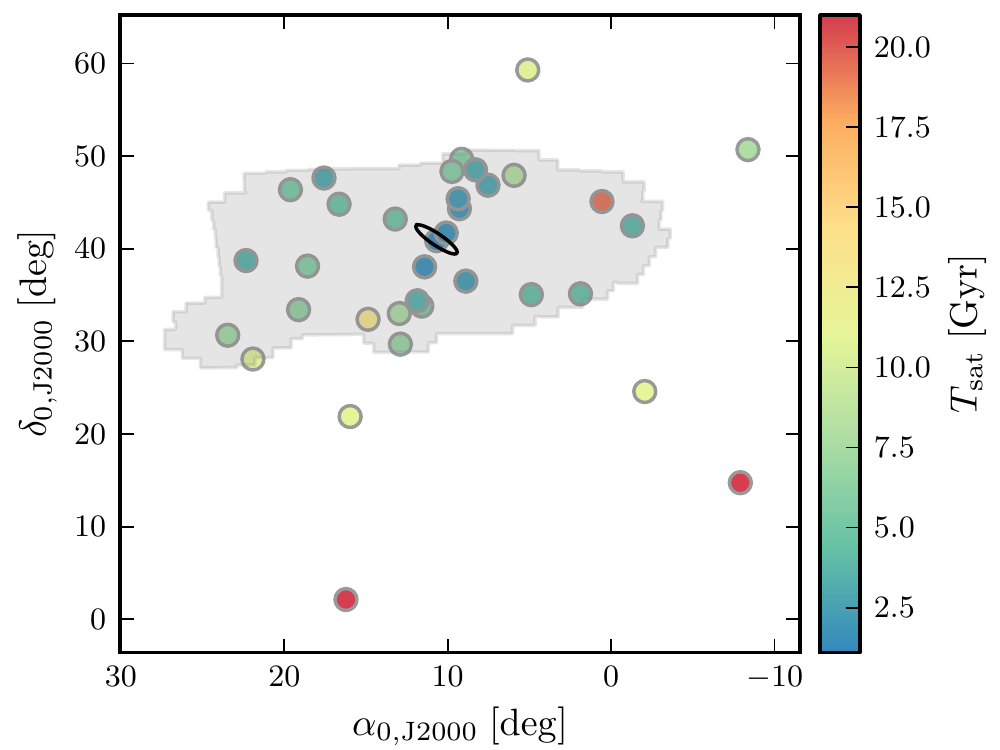}
    \caption{Positions of the satellites in equatorial coordinates; the points are coloured according to the estimated period of the satellite's orbit (see \autoref{sect:orbprops} for details and \autoref{table:orbprops} for full satellite orbital properties).  The ellipse shows the extent of the bright disk of M31.  The PAndAS footprint is shaded to illustrate the bias in the spatial coverage of current deep surveys.}
    \label{fig:sats_ad}
\end{center}
\end{figure}

The position and velocity data for M31 and its satellites are given in \autoref{table:data}.  The locations of the satellites are shown in \autoref{fig:sats_ad} in equatorial coordinates, with the position of M31 shown as an ellipse (the size of which corresponds to the extent of the disk).  The points are coloured according to the orbital periods that we derive in \autoref{sect:orbprops}.  The shaded region shows the PAndAS footprint to illustrate the bias in the spatial coverage of current deep surveys.

Distance estimates for most of the galaxies are from the profiles presented by \citet{conn2012}; although we quote distances and errors in the table, we do not use these directly in the first part of our analysis, instead we use the full heliocentric distance profiles. Where distance profiles are not available, we assume the profiles are Gaussian.

The majority of the dSph velocities come from \citet{tollerud2012} and \citet{collins2012}.  If a satellite has been studied by both groups, we follow the advice of \citet{collins2012} and use the estimate based on a greater number of member stars.  There are some instances in which the heliocentric velocities have asymmetric errors; as these are generally only different by small amounts, we adopt the mean of the error.

We use the M31 proper motions $(\mu_{\alpha}, \mu_{\delta}) = (34.3 \pm 8.6, -20.2 \pm 7.8) \, \mu \mathrm{as \, yr^{-1}}$ recently presented in \citet{vandermarel2012}; that study assumed a fixed M31 distance, whereas here we consider the full posterior distance distribution \citep{conn2012}. This makes it necessary to adjust the proper motions to account for this change.


\section{Monte Carlo simulations}
\label{sect:mc_sims}

With proper motions as well as line-of-sight velocities for all of the M31 satellites, we could determine both the radial and tangential components of the satellite motions. Unfortunately, proper motion data exist for only two of the satellites: IC\,10 and M33; for the rest, there are only line-of-sight velocities.  If we wish to gain insights into the properties of the orbits, we must turn to a more statistical analysis.

To start, we generate a population of $\norb$ satellites around a central host, with their orbits described by the timing argument.  We then view each of the satellites from $\npos$ randomly selected viewing positions in order to determine line-of-sight velocities. Finally, we determine how likely each M31 satellite is in a given model.


\subsection{The timing argument}
\label{sect:ta}

To apply the timing argument to M31 and any one of its satellites, we solve the following set of equations for spherical radius $r$ from the centre of M31, time $t$, radial velocity $\vr$ and tangential velocity $\vt$ of the satellite:
\begin{eqnarray}
    r = a ( 1 - e \ceta ), & \qquad & t = \frac{a}{\lambda} \left( \eta - e
        \seta \right), \nonumber \\
    \vr = \lambda \frac{e \seta}{1 - e \ceta}, & \qquad & \vt = \lambda
        \frac{\sqrt{1 - e^2}}{1 - e \ceta},
    \label{eqn:rtvrvt}
\end{eqnarray}
where $\eta$ is the eccentric anomaly, $e$ is the eccentricity of the orbit, $a$ is the semi-major axis of the orbit, $M$ is the total mass of the system and $\lambda = (GM/a)^{\frac12}$.  For each host-satellite pair, we assume that their mutual gravitational attraction caused them to decouple from the Hubble flow at high redshift. Thus, the time $t$ is the age of the universe, which is $\tu = 13.73^{+0.16}_{-0.15}$ Gyr \citep{spergel2007}.


\subsection{Models}
\label{sect:models}

\begin{figure}
\begin{center}
    \includegraphics[width=\linewidth]{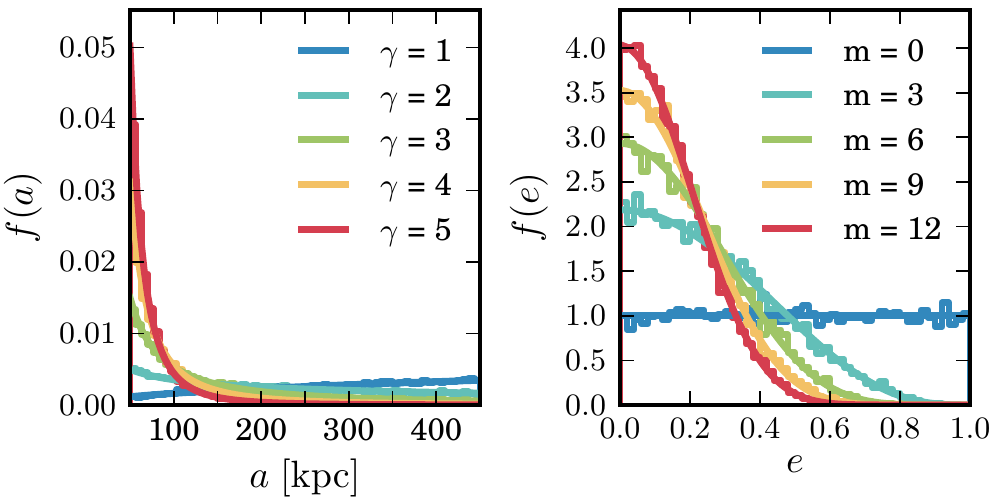}
    \caption{Left panel: the distribution of semi-major axes for different values of $\gamma$.  Right panel: the distribution of orbital eccentricities for different values of $m$.  In both panels, the smooth lines represent the analytical distributions (see \autoref{eqn:fafe}) and the histograms represent samples of selected at random from the given distribution.}
    \label{fig:fafe_dbns}
\end{center}
\end{figure}

In the nearly-Keplerian regime in which the Galactic potential is $\phi \propto r^{-\beta}$ with $\beta \approx 1$, a density distribution that falls like $\rho \propto r^{-\gamma}$ is generated by a phase space distribution function \citep{evans1997}
\begin{eqnarray}
    f(E,L) \propto |L|^{2m} |E|^{\alpha}, & \qquad &
        \alpha = \frac{ (2 - \beta) m + \gamma}{\beta} - \frac{3}{2}
\end{eqnarray}
or equivalently
\begin{equation}
    f( a, e ) \propto ( 1 - e^2 )^m a^{m - \alpha}.
\end{equation}
The distributions of semi-major axes $a$ and eccentricities $e$ are given by
\begin{eqnarray}
    f(a) \propto a^{\frac{3}{2} - \gamma}, & \qquad
        & f(e) \propto ( 1 - e^2 )^m.
    \label{eqn:fafe}
\end{eqnarray}
Note that $\gamma$ alone controls the distribution of semi-major axes whilst $m$ controls the distribution of eccentricities.  The shapes of these distributions are shown in \autoref{fig:fafe_dbns}; the solid lines represent the functional form of the distributions, the histograms show samples selected at random from the given distributions.  The left panel shows the semi-major axis distributions for various $\gamma$; for $\gamma = 1$, the distribution of semi-major axes is nearly flat, with a slight preference for large values of $a$, but as $\gamma$ increases the distributions strongly favour smaller semi-major axes and the distributions are highly peaked. The right panel shows the eccentricity distributions for a range of $m$.  For $m = 0$, the distribution of eccentricities is flat; as $m$ increases, the distributions begin to favour low eccentricities, until for high $m$ the most eccentric orbits are effectively ruled out.

For given $\gamma$ and $m$, we generate $\norb$ $(a,e)$ pairs from these distributions.  We adopt an M31 mass of $\Mand = 1.5 \times 10^{12} \Msun$ \citep{watkins2010} and assume a universal satellite mass of $\Msat = 1 \times 10^7 \Msun$ \citep[e.g.][]{strigari2008}; then the total system mass $M = \Mand + \Msat$.  Now we use \autoref{eqn:rtvrvt} to propagate the orbits forwards until $t = \tu$ (with corresponding eccentric anomaly $\eta = \etau$).  The values of $r$ at this point are the host-satellite separations.

To fix the viewing positions, we select a host-observer distance $d$ from the heliocentric distance profile of M31 \citep{conn2012}.  We generate position angles $(l,b)$, where $l$ is selected at random in $[0,2\pi]$ and $\sin b$ is selected at random in [0,1].  $\npos$ viewing positions are generated separately for each of the $\norb$ orbits.  For each viewing position, the component of the satellite's velocity $\vlos$ along the line-of-sight is calculated.

\begin{figure}
\begin{center}
    \includegraphics[width=\linewidth]{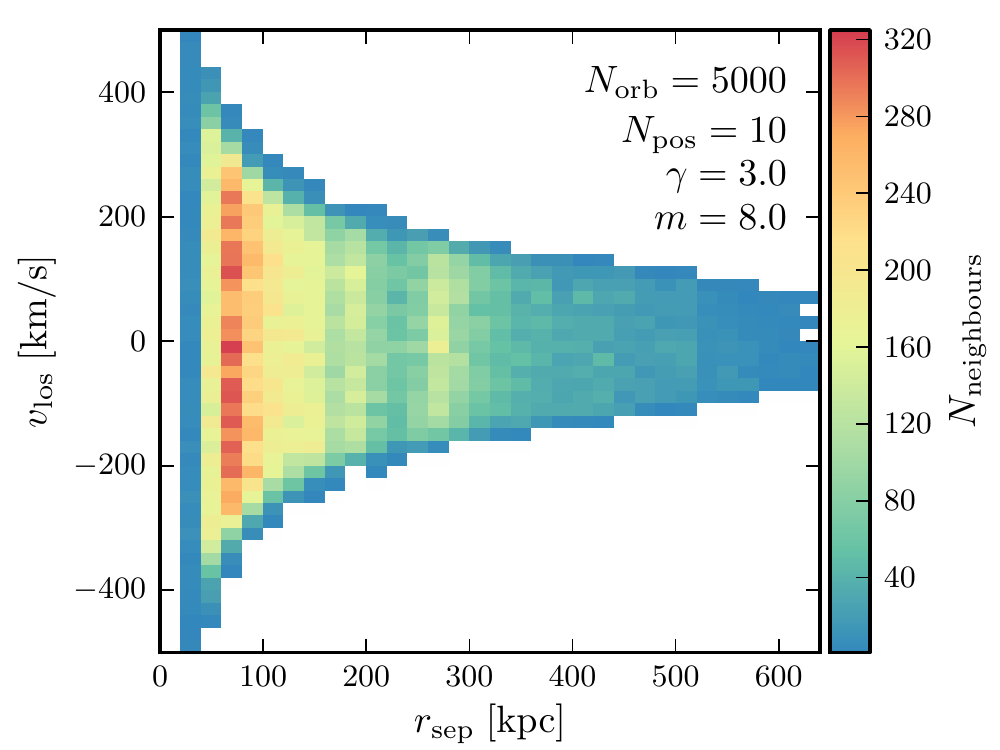}
    \caption{The distribution of host-satellite separations and absolute line-of-sight velocities for a model with $\gamma = 3.0$, $m = 8.0$, $\norb = 5000$ and $\npos = 10$.  The points are binned into pixels of size 20~kpc $\times$ 20~$\kms$ and the pixels coloured according to the number of points therein.  For clarity, the $\vlos$ axis has been truncated at $\pm 500 \; \kms$.}
    \label{fig:mc_rv}
\end{center}
\end{figure}

Now we have a set of $N = \norb \times \npos$ simulated satellites, each with a host-satellite separation distance $r$ and a line-of-sight velocity $\vlos$.  A sample distribution of separation distances and line-of-sight velocities is shown in \autoref{fig:mc_rv}, generated with parameters: $\gamma = 3.0$, $m = 8.0$, $\norb = 5000$ and $\npos = 10$ (see \autoref{sect:setup} for a discussion of how $\norb$ and $\npos$ are chosen).  As a visual aid, to highlight the density of the points, the points have been binned into 10~kpc $\times$ 10~$\kms$ pixels and the pixels coloured according to the number of points they contain.


\subsection{Satellite likelihoods}
\label{sect:satprobs}

For a given model, we wish to determine how likely it is that we observe a particular satellite given the model predictions, and also, how likely it is that we observe the whole satellite system.

For each satellite, we have data $\Thsat$ that consists of the sky coordinates of the satellite (for which we assume no errors), the heliocentric distance of the satellite and errors, the observed radial velocity of the satellite and error, and (where available) the proper motions of the satellite and errors; together with the same data for M31.  We assume the error distributions are Gaussian, although for the satellite distances, we allow the positive error to be different from the negative error.

From this data, we can calculate the distributions of separations $r$ of the satellite from the centre of M31 and the line-of-sight velocities $\vlos$ of the satellite corrected for solar peculiar motion\footnote{$\left( U_0, V_0, W_0 \right) = ( 10.00 \pm 0.36, 5.25 \pm 0.62, 7.17 \pm 0.38 ) \kms$ \citep{dehnen1998}.}, motion of the Local Standard of Rest (LSR)\footnote{$V_{\mathrm{LSR}} = 220 \pm 20 \kms$ \citep{kerr1986}.} and the relative motion of the Milky Way and M31.  In what follows, for greater clarity, we will drop the \textsc{los} subscript on $\vlos$, but we will always be considering line-of-sight velocities.

We start by generating $10^5$ sets of data $\Thsat$ from their respective distributions, and calculate $r$ and $v$ for each set.  As expected, the distribution of velocities is approximately Gaussian, with a mean $\vsat$ similar to that calculated using the data values and a standard deviation $\evsat$ comparable to that inferred from standard error propagation.  The distribution of separations requires some further thought.  The separation distance $r$ is calculated via,
\begin{equation}
    r^2 = d^2 + \dand^2 - 2 d \dand \cos \theta,
    \label{eqn:rsep}
\end{equation}
where $d$ is the heliocentric distance of the satellite, $\dand$ is the heliocentric distance of M31 and $\theta$ is their angular separation.  There is a minimum separation distance $\rmin = \dand \sin \theta$ (achieved when $d = \dand \cos \theta$).  Even when the satellite heliocentric distances are normally distributed, the resulting separation distribution cannot be obtained by simply truncating the Gaussian at $\rmin$ and renormalising because the distribution can become significantly non-Gaussian.  The distribution accumulates near the minimum because of the complicated dependence of the separation distance on the satellite position.  When the satellite distance profiles are non-Gaussian \citep[as they are for many satellites in this analysis, see][]{conn2012}, the separation distributions become yet more complex.
\begin{figure*}
\begin{center}
   \includegraphics[width=0.33\linewidth]{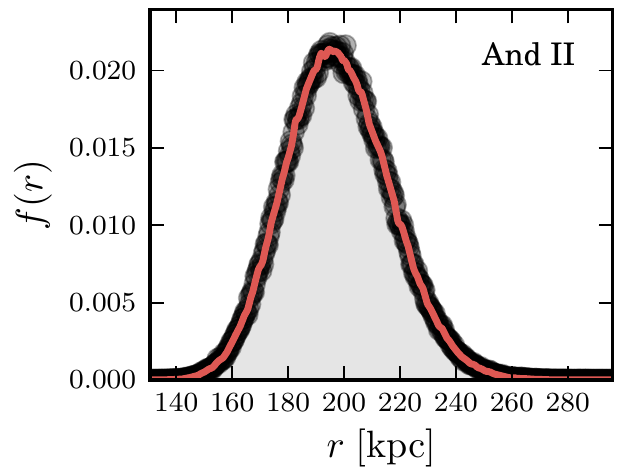}
   \includegraphics[width=0.33\linewidth]{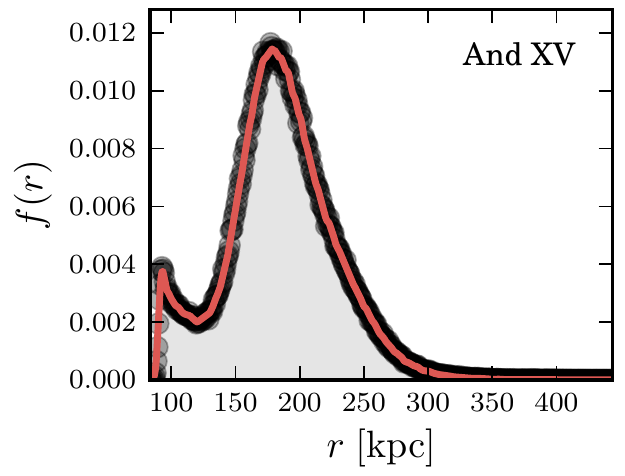}
   \includegraphics[width=0.33\linewidth]{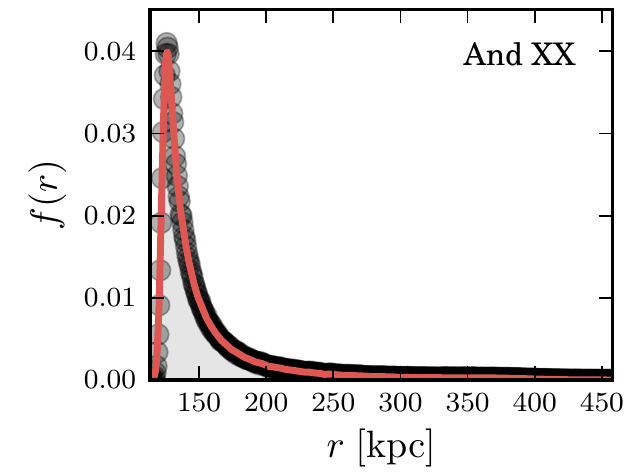}
   \caption{The distribution of separations for And\,II (left), And\,XV (middle) and And\,XX (right).  And\,II is sufficiently distant with small enough errors so that the shape of the separation profile is similar to that of the heliocentric distance profile.  And\,XV is an intermediate case, as its separation distribution is similar to the heliocentric distance distribution at larger separations, but this breaks down close to the minimum separation.  For And\,XX, the distribution finds a minimum separation distance around which the distribution becomes highly peaked.  The red lines show the functions fitted to the distributions.}
   \label{fig:pr}
\end{center}
\end{figure*}

\autoref{fig:pr} shows the distribution of separations for three typical satellites.  Satellites such as And\,II are either far enough away from the centre of M31 or have small enough errors that the minimum separation is never reached and the distribution does not pile up at small separations.  For satellites such as And\,XX, the distributions are highly distorted with a sharp peak at or near the minimum separation distance.  The final example, And\,XV, is an intermediate case that is just starting to show an increase in separations near the minimum separation distance.  In order to account for these profiles in our models, we approximate the functional forms of the likelihood distributions by fitting splines to the smoothed histograms.  These functions $f(r)$ are shown by the red lines in \autoref{fig:pr}.  We include the full set of separation-distance distributions in \autoref{fig:pr_full}.

For each model, we have a set of $N$ pairs of separation distances and line-of-sight velocities $\left\{ \Thi = (\ri,\vi) \right\}_{i=1}^N$. For each point, we ask what is the likelihood of observing $\Thsat$,
\begin{eqnarray}
    P \left( \Thsat | \Thi, \gamma, m \right) & = & C P \left( \Thsat | \ri,
        \gamma, m \right) P \left( \Thsat | \vi, \gamma, m \right) \nonumber \\
    & = & C f( \ri ) \exp \left( -\frac{\left( \vi - \vsat \right)^2}
        {2 \evsat^2} \right),
\end{eqnarray}
for normalising constant C.  The likelihood of observing the satellite given the model is then the weighted sum over all points $\Thi$,
\begin{equation}
    P \left( \Thsat | \gamma, m \right) = \sum_{i=1}^N P \left( \Thsat | \Thi,
        \gamma, m \right) P \left( \Thi \right).
\end{equation}
The weights $P(\Thi) = 1/N$ in this case, as the priors on $\Thi$ are already accounted for in the Monte Carlo modelling.

The likelihood of observing the whole satellite system $\Thsys$, is the product of the individual likelihoods for each satellite and is thus
\begin{equation}
    P \left( \Thsys | \gamma, m \right) = \prod_{\mathrm{satellites}}
        P \left( \Thsat | \gamma, m \right).
\end{equation}


\section{Results}
\label{sect:results}


\subsection{Setup}
\label{sect:setup}

Recall, we wish to sample $\norb$ orbits from the distributions given in \autoref{eqn:fafe}, and to ``observe" each orbit from $\npos$ viewing positions.  The total number of points $\ntot = \norb \times \npos$ should be as large as possible to ensure that the $rv$-space is well sampled. We find that $\ntot = 10^6$ is sufficient for this purpose.  $\norb$ controls the number of radii sampled, while $\npos$ is the number of velocities sampled at each radius.  Provided $\norb$ is high enough, lowish values of $\npos$ are acceptable as the radial distribution is sampled well enough to give a representative set of radii and velocities. We find that using $\norb = 10^5$ and $\npos = 10$ enables us to well sample the $rv$-space while keeping computing times down. The range of eccentricities sampled is $e \in [0, 1]$. For the semi-major axes, we have to take care.  The timing argument assumptions break down very near to the centre of M31, so we do not allow $a$ to become too small. To determine the range of semi-major axes, we take the 6.4 and 93.6 (1.5\,$\sigma$) percentiles of the combined separation-distance distributions, rounded to the nearest 50~kpc, to give $a \in [50,450]$~kpc.


\subsection{Satellite system}
\label{sect:system}

\begin{figure}
\begin{center}
    \includegraphics[width=\linewidth]{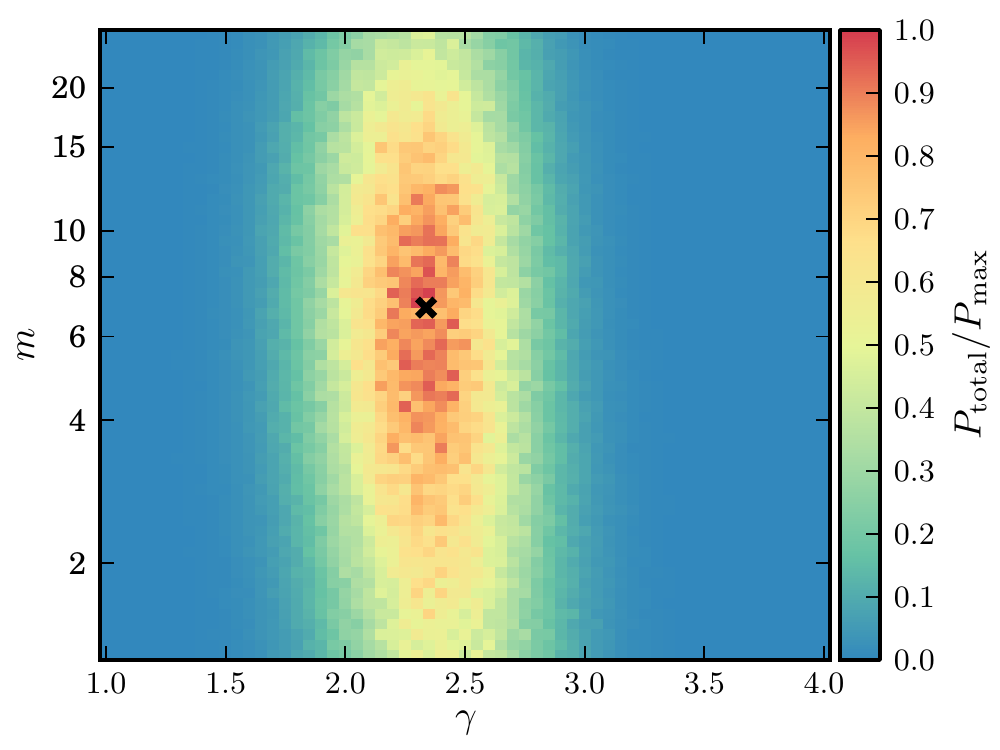}
    \caption{The likelihood of observing the M31 satellite system as a function of $\gamma$ (which controls the distribution of semi-major axes) and $m$ (which controls the distribution of orbital eccentricities).  The black cross represents the most likely model as determined by marginalising over $\gamma$ and $m$ in turn.}
    \label{fig:ptot}
\end{center}
\end{figure}

We search an 61x61 grid of models with $\gamma \in [1,4]$ and $\ln m \in [0.25,3.25]$, which corresponds to $m \in [1.3,25.8]$.  The sampling is performed logarithmically in $m$ instead of linearly. The right panel of \autoref{fig:fafe_dbns} shows that a change in $m$ for small values of $m$ has a more significant effect on the shape of the eccentricity distribution than for large values of $m$. 

The total likelihood of observing the whole M31 satellite system is shown in \autoref{fig:ptot} as a function of $\gamma$ and $m$.  The value of $\gamma$ is well constrained in the range that we have considered indicating that there is a very specific distribution of semi-major axes that fit the observed satellite distribution.  The distribution of eccentricities is less-well constrained; it is for this reason that we consider such a large range of $m$ values.
\begin{figure}
\begin{center}
    \includegraphics[width=\linewidth]{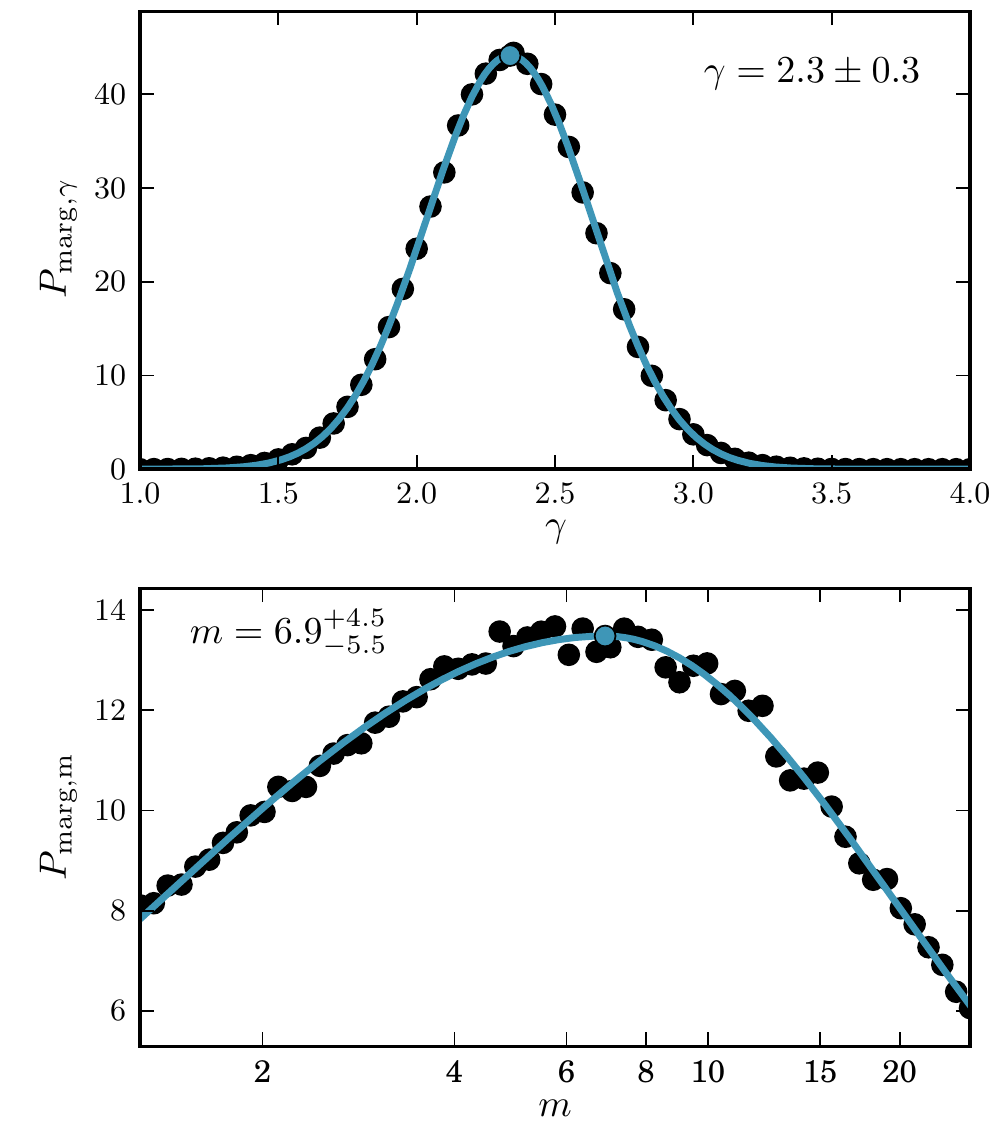}
    \caption{Marginalised likelihood of the M31 satellite system as a
      function of $\gamma$ (left) and $m$ (right).  $\gamma$ is
      tightly constrained at $\gamma = 2.5 \pm 0.3$.  $m$ shows more
      scatter but still reaches a clear peak at $m =
      7.4_{-8.4}^{+9.7}$.}
    \label{fig:pmarg}
\end{center}
\end{figure}
There is some scatter between models as $\ntot$ is finite.  To determine the values of $\gamma$ and $m$ that maximise the likelihood of the satellite system, we marginalise over $m$ and $\gamma$ in turn. The results are shown in \autoref{fig:pmarg}, with an asymmetric Gaussian fitted to the marginalised points to determine the peak of the distribution.  The top panel shows the marginalised likelihood as a function of $\gamma$ and it is again clear that $\gamma$ is well constrained.  The bottom panel shows the marginalised likelihood as a function of $m$; the constraints are weaker here but the distribution does reach a clear peak.
\begin{figure}
\begin{center}
    \includegraphics[width=\linewidth]{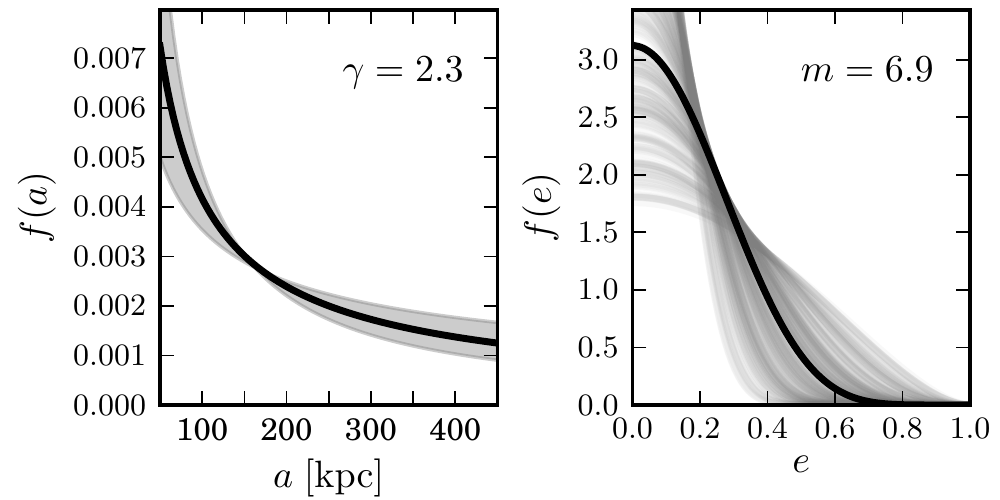}
    \caption{The distributions of semi-major axes and orbital eccentricity for the values of $\gamma$ and $m$ that best reproduce the M31 satellite system.}
    \label{fig:fafebest}
\end{center}
\end{figure}

The best model, after marginalisation, is found to have $\gamma = 2.3 \pm 0.3$ and $m = 6.9_{-5.5}^{+4.5}$.  This is marked with a black cross in \autoref{fig:ptot}.  The corresponding orbital distribution functions are shown in \autoref{fig:fafebest}.  From these, we can see that such a value for $\gamma$ favours short semi-major axes, while still allowing a small number of large semi-major axis lengths (which are required for the outermost satellites).  Such values for $m$ strongly favour low eccentricities, and rule out higher eccentricities.


\subsection{Individual satellites}
\label{sect:sats}

\begin{figure*}
\begin{center}
    \includegraphics[width=\linewidth]{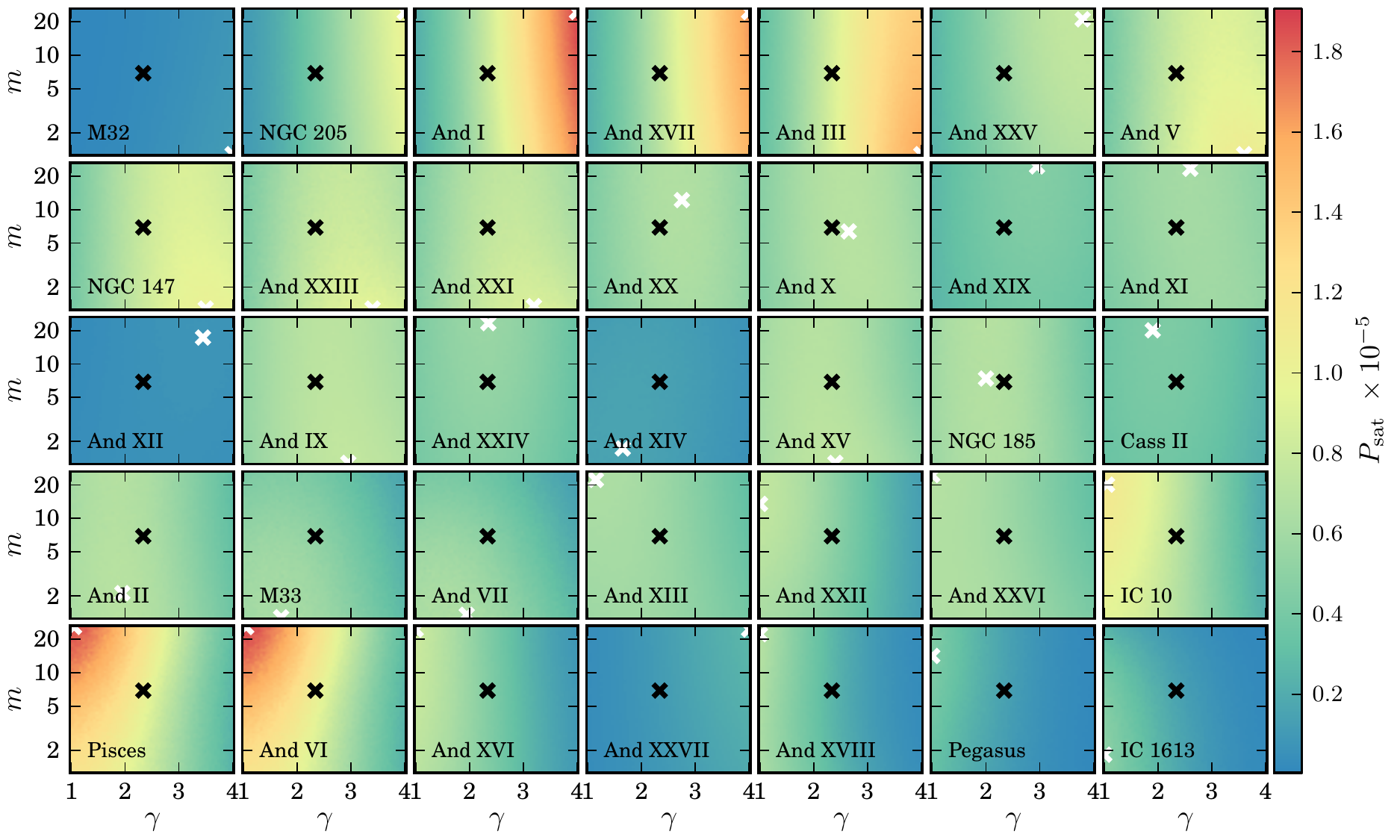}
    \caption{The likelihood of observing each M31 satellite as a function of the model parameters $\gamma$ and $m$ (which determine the distribution of semi-major axes and eccentricities respectively).  The colours of the pixels indicate with likelihood, with all satellites plotted using the same scale to  allow easy comparison.  The satellites are ordered by the median of their separation distribution.  Both And\,XII and And\,XIV have a very low likelihood for all models, despite having separations similar to the bulk of the satellite population, suggesting that they are very unusual objects.}
    \label{fig:psat}
\end{center}
\end{figure*}

Now we turn to the likelihoods of the individual satellites; these are shown as a function of $\gamma$ and $m$ in \autoref{fig:psat} with the satellites ordered according to the median of their separation distance distribution.  The colours of the pixels represent the likelihood of the satellite in a given model; all the satellites are shown on the same likelihood (colour) scale to aid comparison.  The best model for the entire satellite system is shown by the black cross in each panel.  The best model for each individual satellite is shown as a white cross.

M32 appears to be unlikely for all of the models we have considered here, which is probably due to its proximity to M31; the example model shown in \autoref{fig:mc_rv} clearly shows that very small separation distances are hard to achieve with the range of semi-major axes we have used.  And\,XVIII, Pegasus and IC\,1613 also appear to be rather unlikely, but in their case it is due to their large separation distances.  And\,XXVII also has a low-likelihood. The large errors on its heliocentric distance (which arise because it is in the process of being tidally disrupted and may even be gravitationally unbound \citep{collins2012}) lead to a correspondingly large range of separation distances (see \autoref{fig:pr_full}).

And\,XII and And\,XIV also have low likelihoods for all values of $\gamma$ and $m$.  We cannot make the same arguments for these satellites, as both are found at distances similar to the main bulk of the satellite population. It is their large line-of-sight velocities that make them unusual.  This indicates that they are not well described by the models that we have been considering here.  To apply the timing argument, we have made two assumptions: that the orbits are in the nearly-Keplerian regime; and that the Milky Way and other satellites make only a negligible contribution to the orbits.  In the Monte Carlo simulations, we have made a third assumption that the satellites are part of a well-mixed population with orbits consistent with the distributions functions described in \autoref{eqn:fafe}.  We believe that the first assumption does hold good at these intermediate radii, and, in \autoref{sect:introduction}, we discussed our reasons for the second assumption in some detail.  Thus, it is the third assumption that we suspect is invalid for these two satellites and we conclude that they are not part of a well-mixed population.

We observe a trend in the preferred value of $\gamma$ with separations; the innermost satellites all favour high values of $\gamma$, that is, they require very small semi-major axes.  Moving out in radius, the preferred value of $\gamma$ shifts to lower values, allowing for increasing numbers of satellites with larger semi-major axes.  For the outermost satellites, small values of $\gamma$ are strongly preferred as these models necessarily allow orbits with large semi-major axes.  The exception to this trend is And\,XXVII, but it has very large distance errors and should be treated with caution.

\begin{figure}
\begin{center}
    \includegraphics[width=\linewidth]{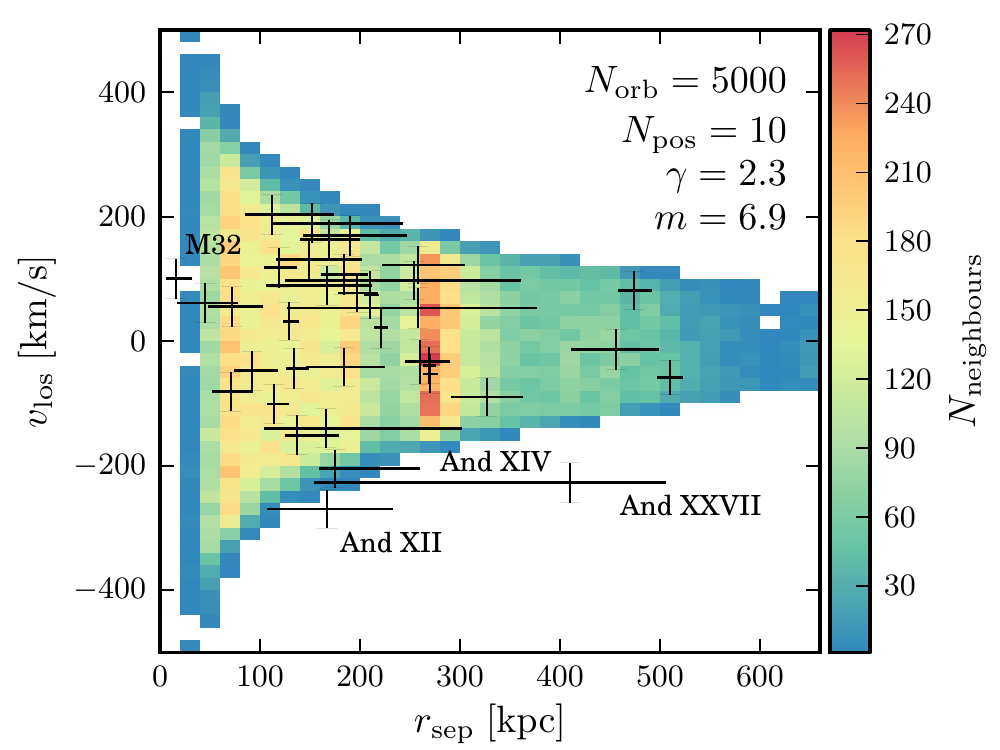}
    \caption{A sample of separations and line-of-sight velocities generated from the best model (for details see \autoref{fig:mc_rv}), with the observed values (and errors) for the satellites overplotted.  And\,XII and And\,XIV are outliers to the model distribution, explaining their low likelihoods.  M32 is also an outlier as there are few model orbits found at such low separations.  And\,XXVII is also marked to highlight the very large uncertainty on its separation.}
    \label{fig:rvbest}
\end{center}
\end{figure}

A sample of separations and line-of-sight velocities generated from the best model is shown in \autoref{fig:rvbest}.  Just as in \autoref{fig:mc_rv}, the points are coloured according to their number of nearest neighbours as an indication of the density.  The positions and velocities of the satellites are also shown, with And\,XII, And\,XIV, M32 and And\,XXVII marked.  As we have already noted, M32 is very close in to M31 and the model has trouble producing satellites at such small distances.  The very large uncertainty on the distance of And\,XXVII is also apparent here.  And\,XII and And\,XIV are very much outliers to the distribution of model points, which explains the low likelihoods that we recover.


\subsection{Orbital properties}
\label{sect:orbprops}

\begin{figure*}
\begin{center}
    \includegraphics[width=\linewidth]{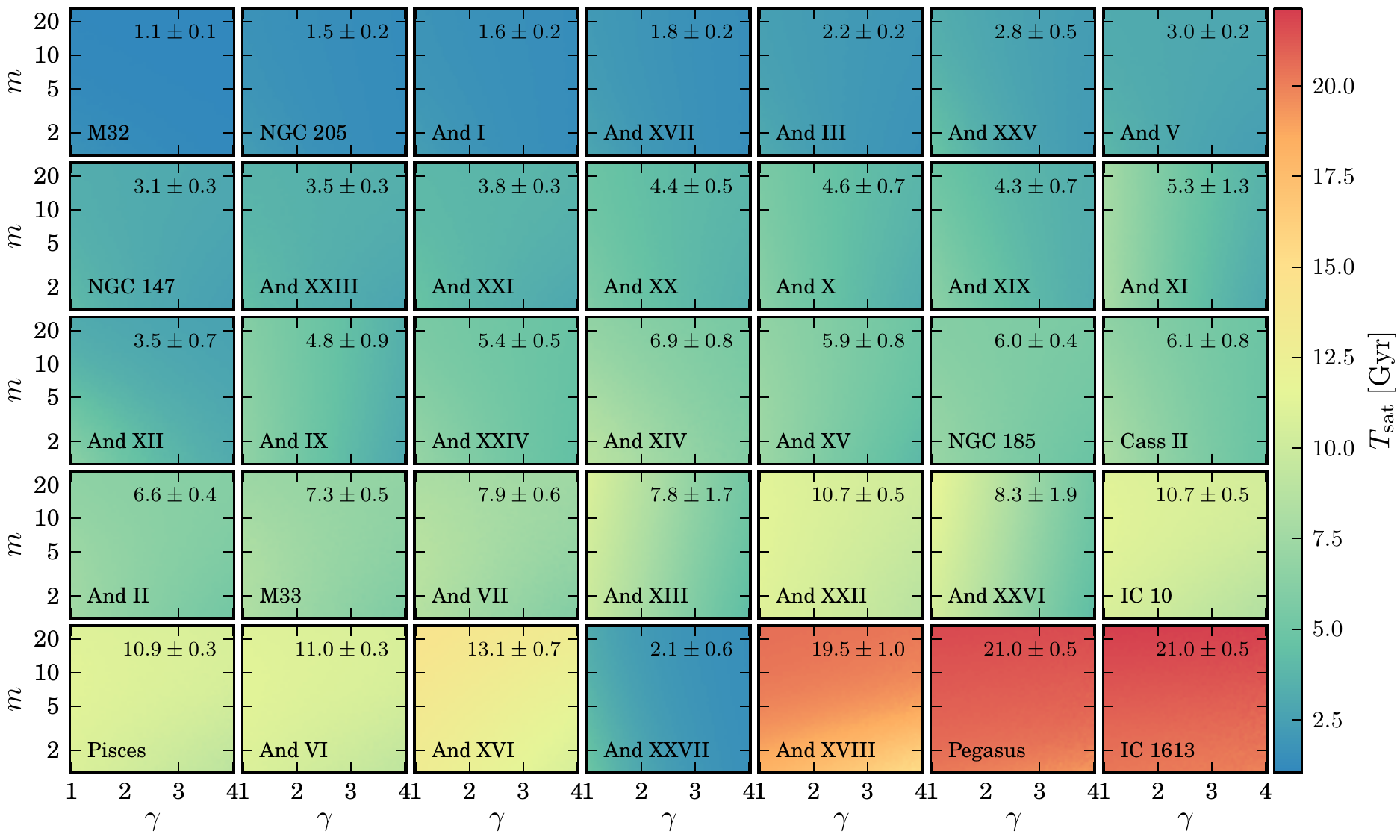}
    \caption{Weighted-mean orbital period as a function of model parameters $\gamma$ and $m$ for each satellite.  The colours of the pixels indicate the mean period.  The periods change significantly from satellite to satellite, generally increasing with separation. However the periods change little over the range of models considering, indicating that the mean period is a robust measure for each satellite.  For each satellite, the mean period for all the models is shown in the top-right corner of each panel.}
    \label{fig:pdsat}
\end{center}
\end{figure*}

For each orbit $i$ in a given model, we have a semi-major axis $a$ and eccentricity $e$. From these, we can calculate the period $T$, apocentric distance $\ra$ and pericentric distance $\rp$.  Then for orbit parameter $X$, the weighted-mean period $\overline{X}_{\mathrm{sat}}$ of the satellite for that model is given by
\begin{equation}
    \overline{X}_{\mathrm{sat}} \left( \gamma, m \right ) = \frac{\sum_{i=1}^N P
        \left( \Thsat | \Thi, \gamma, m \right) X_{\mathrm{i}}}
        {\sum_{i=1}^N P \left( \Thsat | \Thi, \gamma, m \right)}.
\end{equation}
The distributions of mean periods $\overline{T}_{\mathrm{sat}}$ are shown in \autoref{fig:pdsat}.  The mean period for all the models is shown in the upper-right corner of each plot.  For all satellites, the average period changes very little over the range of models considered here.  As expected, the periods of the satellite become larger as their separation from M31 increases.  And\,XXVII is somewhat anomalous in having a very small period, whereas the satellites at similar distances have very long periods, but this is likely due to the fact that the errors on its heliocentric distance are large.

And\,XII and And\,XIV seem to have average periods consistent with those of their neighbours.  The period of And\,XII is estimated at $\overline{T} \sim 3.5$~Gyr and the period of And\,XIV is estimates at $\overline{T} \sim 7$~Gyr.  Thus, for these models, And\,XIV is predicted to be on its second or third orbit, and And\,XII on its fourth; i.e. neither satellite is on its first infall into the M31 system.  However, from \autoref{fig:psat}, we conclude that these satellites are unlikely for all of the models that we have considered, so it is unwise to put too much faith into these periods.

\begin{table*}
    \caption{Mean orbital properties for the M31 satellites.}
    \label{table:orbprops}
    \begin{tabular}{lccccc}
        \hline
        \hline
        & period & semi-major axis
            & eccentricity & apocentre
            & pericentre \\
        & (Gyr) & (kpc) & & (kpc) & (kpc) \\
        \hline
        And\,I & 1.6 $\pm$ 0.2 & 74 $\pm$ 4 & 0.21 $\pm$ 0.07 & 90 $\pm$ 8 & 59 $\pm$ 5 \\
        And\,II & 6.6 $\pm$ 0.4 & 192 $\pm$ 8 & 0.23 $\pm$ 0.08 & 235 $\pm$ 17 & 148 $\pm$ 17 \\
        And\,III & 2.2 $\pm$ 0.2 & 90 $\pm$ 6 & 0.23 $\pm$ 0.08 & 111 $\pm$ 10 & 70 $\pm$ 9 \\
        And\,V & 3.0 $\pm$ 0.2 & 114 $\pm$ 6 & 0.23 $\pm$ 0.08 & 140 $\pm$ 9 & 88 $\pm$ 11 \\
        And\,VI & 11.0 $\pm$ 0.3 & 273 $\pm$ 6 & 0.17 $\pm$ 0.05 & 320 $\pm$ 10 & 226 $\pm$ 18 \\
        And\,VII & 7.9 $\pm$ 0.6 & 217 $\pm$ 11 & 0.28 $\pm$ 0.09 & 277 $\pm$ 19 & 157 $\pm$ 22 \\
        And\,IX & 4.8 $\pm$ 0.9 & 152 $\pm$ 20 & 0.23 $\pm$ 0.08 & 186 $\pm$ 26 & 118 $\pm$ 20 \\
        And\,X & 4.6 $\pm$ 0.7 & 149 $\pm$ 14 & 0.22 $\pm$ 0.07 & 181 $\pm$ 19 & 116 $\pm$ 16 \\
        And\,XI & 5.3 $\pm$ 1.3 & 159 $\pm$ 26 & 0.22 $\pm$ 0.07 & 193 $\pm$ 32 & 125 $\pm$ 25 \\
        $\ast$\,And\,XII & 3.5 $\pm$ 0.7 & 125 $\pm$ 15 & 0.21 $\pm$ 0.07 & 152 $\pm$ 27 & 98 $\pm$ 5 \\
        And\,XIII & 7.8 $\pm$ 1.7 & 208 $\pm$ 32 & 0.22 $\pm$ 0.08 & 253 $\pm$ 37 & 163 $\pm$ 33 \\
        $\ast$\,And\,XIV & 6.9 $\pm$ 0.8 & 197 $\pm$ 14 & 0.23 $\pm$ 0.06 & 243 $\pm$ 28 & 151 $\pm$ 9 \\
        And\,XV & 5.9 $\pm$ 0.8 & 177 $\pm$ 15 & 0.23 $\pm$ 0.08 & 217 $\pm$ 21 & 136 $\pm$ 21 \\
        And\,XVI & 13.1 $\pm$ 0.7 & 305 $\pm$ 11 & 0.22 $\pm$ 0.07 & 371 $\pm$ 13 & 240 $\pm$ 29 \\
        And\,XVII & 1.8 $\pm$ 0.2 & 79 $\pm$ 6 & 0.21 $\pm$ 0.07 & 95 $\pm$ 10 & 62 $\pm$ 7 \\
        $\ast$\,And\,XVIII & 19.5 $\pm$ 1.0 & 400 $\pm$ 15 & 0.22 $\pm$ 0.08 & 487 $\pm$ 16 & 313 $\pm$ 44 \\
        And\,XIX & 4.3 $\pm$ 0.7 & 141 $\pm$ 15 & 0.20 $\pm$ 0.06 & 170 $\pm$ 22 & 112 $\pm$ 12 \\
        And\,XX & 4.4 $\pm$ 0.5 & 145 $\pm$ 10 & 0.21 $\pm$ 0.07 & 176 $\pm$ 15 & 114 $\pm$ 12 \\
        And\,XXI & 3.8 $\pm$ 0.3 & 133 $\pm$ 7 & 0.24 $\pm$ 0.08 & 165 $\pm$ 11 & 102 $\pm$ 14 \\
        And\,XXII & 10.7 $\pm$ 0.5 & 267 $\pm$ 9 & 0.20 $\pm$ 0.06 & 322 $\pm$ 14 & 213 $\pm$ 20 \\
        And\,XXIII & 3.5 $\pm$ 0.3 & 127 $\pm$ 6 & 0.24 $\pm$ 0.09 & 157 $\pm$ 10 & 97 $\pm$ 13 \\
        And\,XXIV & 5.4 $\pm$ 0.5 & 167 $\pm$ 10 & 0.20 $\pm$ 0.06 & 202 $\pm$ 18 & 133 $\pm$ 12 \\
        And\,XXV & 2.8 $\pm$ 0.5 & 108 $\pm$ 11 & 0.20 $\pm$ 0.06 & 130 $\pm$ 16 & 85 $\pm$ 10 \\
        And\,XXVI & 8.3 $\pm$ 1.9 & 216 $\pm$ 36 & 0.23 $\pm$ 0.08 & 265 $\pm$ 43 & 168 $\pm$ 38 \\
        $\ast$\,And\,XXVII & 2.1 $\pm$ 0.6 & 85 $\pm$ 14 & 0.19 $\pm$ 0.06 & 101 $\pm$ 19 & 69 $\pm$ 10 \\
        Cass\,II & 6.1 $\pm$ 0.8 & 181 $\pm$ 15 & 0.21 $\pm$ 0.06 & 218 $\pm$ 24 & 144 $\pm$ 14 \\
        IC\,10 & 10.7 $\pm$ 0.5 & 268 $\pm$ 9 & 0.21 $\pm$ 0.07 & 323 $\pm$ 12 & 212 $\pm$ 24 \\
        $\ast$\,IC\,1613 & 21.0 $\pm$ 0.5 & 422 $\pm$ 7 & 0.31 $\pm$ 0.06 & 552 $\pm$ 15 & 293 $\pm$ 29 \\
        $\ast$\,M32 & 1.1 $\pm$ 0.1 & 60 $\pm$ 3 & 0.27 $\pm$ 0.10 & 77 $\pm$ 9 & 44 $\pm$ 4 \\
        M33 & 7.3 $\pm$ 0.5 & 206 $\pm$ 9 & 0.27 $\pm$ 0.08 & 261 $\pm$ 19 & 151 $\pm$ 18 \\
        NGC\,147 & 3.1 $\pm$ 0.3 & 117 $\pm$ 6 & 0.23 $\pm$ 0.08 & 144 $\pm$ 10 & 90 $\pm$ 11 \\
        NGC\,185 & 6.0 $\pm$ 0.4 & 180 $\pm$ 8 & 0.22 $\pm$ 0.07 & 220 $\pm$ 16 & 141 $\pm$ 15 \\
        NGC\,205 & 1.5 $\pm$ 0.2 & 72 $\pm$ 4 & 0.21 $\pm$ 0.07 & 87 $\pm$ 8 & 57 $\pm$ 5 \\
        $\ast$\,Pegasus & 21.0 $\pm$ 0.5 & 423 $\pm$ 6 & 0.19 $\pm$ 0.05 & 504 $\pm$ 12 & 342 $\pm$ 24 \\
        Pisces & 10.9 $\pm$ 0.3 & 272 $\pm$ 6 & 0.18 $\pm$ 0.05 & 319 $\pm$ 10 & 224 $\pm$ 18 \\
        \hline
        \hline
    \end{tabular}
    
    \raggedright
    \medskip
    \textbf{Notes}: * Objects are unlikely to be correctly described by the class of models used to generate these orbital properties and should be used with caution. \\
\end{table*}

We can go one step further and consider the mean orbital parameters of each satellite over all models that we have considered.  To do this we take the weighted mean of the weighted-mean orbital parameter, using the model likelihoods for each satellite as the weights.  For orbital parameter $X$, that is,
\begin{equation}
    \left< \overline{X}_{\mathrm{sat}} \right> = \frac{\sum_{\gamma, m} P
        \left( \Thsat | \gamma, m \right) \overline{X}_{\mathrm{sat}} \left(
        \gamma, m \right ) } {\sum_{\gamma, m} P \left( \Thsat | \gamma, m
        \right)}.
\end{equation}
Similarly, we calculate the weighted variance of the parameters to estimate the uncertainty on these values.  We do this for each of the period, semi-major axis, eccentricity, apocentre and pericentre and for each satellite; the results are displayed in \autoref{table:orbprops}.

A number of orbits can reproduce a given separation distance and, whilst their semi-major axes are well constrained, their eccentricities can show more variation.  Nonetheless, the weighted means of the eccentricities reported in \autoref{table:orbprops} are all very similar, and are comparable to the means of the theoretical distributions in \autoref{eqn:fafe}. Thus, this is a consequence of the assumption that the satellites are well-mixed and have a power-law density profile. If this assumption were dropped -- for example, if the profile is broken -- then the eccentricities would show greater variation.

The range of semi-major axes sampled by the distribution functions is $[50,450]$~kpc; we discussed the motivations for this choice in \autoref{sect:setup}.  Suppose we consider curtailing this range at the upper end and re-run our simulations over a coarse grid using a range $[50,350]$~kpc.  The innermost satellites are oblivious to this change.  The outermost satellites (And\,XVIII, Pegasus and IC\,1613) are significantly affected.  For example, IC\,1613 lies at a distance of $510 \pm 12$~kpc from the centre of M31.  Orbits with semi-major axes $a \sim 450$~kpc need only be mildly eccentric to produce a satellite at this distance.  Orbits with semi-major axes $a \sim 350$~kpc need to be more eccentric to reach out to $\sim$500 kpc.  So distribution functions with lower $m$ are now preferred that permit more orbits with larger eccentricities.  The predicted weighted means of the eccentricities for the outermost satellites do change significantly, namely $\overline{e} \sim 0.60$ for And\,XVIII, $\overline{e} \sim 0.87$ for Pegasus and $\overline{e} \sim 0.95$ for IC\,1613.


\section{Infalling groups}
\label{sect:infall}

Using simulations of galaxy formation in a hierarchical merging cosmogony, \citet{li2008b} found that nearly a third of the dark matter subhalos were accreted in groups.  Such behaviour is not unexpected. \citet{sales2007} postulate that outlying satellites on extreme orbits may have been part of an infalling pair. When the pair interact with the host, the heavier member can maintain a bound orbit, but the lighter satellite is kicked out onto a high-energy orbit. Observations also show ample evidence for group infall; \citet{dejong2010} recently presented evidence that Leo IV and Leo V could be a bound ``tumbling pair" falling into the Milky Way and, in their discovery of And\,XXII, \citet{martin2009} noted that it lies closer in projection to M33 than M31 and suggested that it could be a satellite of M33.  With orbital properties for all of the M31 satellites, it is worth taking a moment to consider whether there is any evidence that groups of satellites can be associated and may have had a common origin.

First we consider whether And\,XXII ($a = 267 \pm 9$~kpc, $e = 0.20 \pm 0.06$) might be not a satellite of M31 but of M33, which itself has parameters $a = 206 \pm 9$~kpc, $e = 0.27 \pm 0.08$. \citet{tollerud2012} and \citet{chapman2012} both found radial velocities consistent with this hypothesis. The latter authors analysed N-body simulations of the M31-M33 system, which suggest that And\,XXII may indeed be a satellite of M33. We find that the semi-major axes are not similar, from which we conclude that the two satellites are not moving independently along the same orbit; but whether this means they are orbiting each other or are unrelated, we cannot determine from this analysis.

NGC\,147 ($a = 117 \pm 6$~kpc, $e = 0.23 \pm 0.08$) and NGC\,185 ($a = 180 \pm 8$~kpc, $e = 0.22 \pm 0.07$) are very close together on the sky and a connection between them has long been suspected.  However the orbits we find for these two objects in our study do not support such a conclusion. Their eccentricities are similar, but their semi-major axes are quite different, so much so that NGC\,147 completes two orbits in the time that NGC\,185 takes to complete one.  Recent discovery Cass\,II ($a = 181 \pm 15$~kpc, $e = 0.21 \pm 6$) lies close to this pair and could be a third member of their subsystem, if such a subsystem exists \citep[see e.g.,][]{irwin2012}. \citet{collins2012} argue that a connection is possible based on the line-of-sight velocity they measure for Cass\,II.  Our results certainly favour a connection between Cass\,II and NGC\,185 as the orbital parameters we find for them are in close agreement, but not for Cass\,II and NGC\,147.

To find other possible infalling groups, we select satellites with semi-major-axis and eccentricity estimates that overlap within their errors. This gives us 47 pairs of satellites.  We further require that the pairs be in close proximity, namely less than 100~kpc apart.  This corresponds to an angle of 7.4$^\circ$ at the distance of M31, so we keep only pairs with separations smaller than this value.  This leaves us with 14 pairs (of which eight are individual pairs and the rest form two groups of three).  Finally, we calculate separation profiles for each of these pairs and calculate the median separation distance. We reject all pairs with median separation larger than 100~kpc, to obtain three candidate pairs and one candidate group, of which the Cass\,II - NGC\,185 pair (separation $46^{+32}_{-29}$~kpc) has already been discussed.  The other two pairs are: And\,I ($a = 74 \pm 4$~kpc, $e = 0.21 \pm 0.07$) and And\,XVII ($a = 79 \pm 6$~kpc, $e = 0.21 \ 0.07$), which have a separation $84^{+9}_{-3}$~kpc; and And\,IX ($a = 152 \pm 20$~kpc, $e = 0.23 \pm 0.08$) and And\,X ($a = 149 \pm 14$~kpc, $e = 0.22 \pm 0.07$), which have a separation $61^{+38}_{-24}$~kpc.  The group is composed of And\,V ($a = 114 \pm 6$~kpc, $e = 0.23 \pm 0.08$), And\,XXV ($a = 108 \pm 11$~kpc, $e = 0.20 \pm 0.06$) and NGC\,147; the And\,V-And\,XXV separation is $95^{+42}_{-9}$~kpc, the And\,V-NGC\,147 separation is $84^{+5}_{-4}$~kpc and the And\,XXV-NGC\,147 separation is $41^{+43}_{-14}$~kpc.  Further study is required to determine whether these candidate groups are real.


\section{Minimum-mass solutions}
\label{sect:minmass}

In the modelling, we assumed that all of the satellites are part of a well-mixed population described by the distribution functions in \autoref{eqn:fafe}. We believe it is this assumption that is not appropriate for And\,XII and And\,XIV. Here, we relax that assumption and then use the timing argument to analyse the orbits of these two satellites, placing no priors on the orbital properties.


\subsection{The timing argument revisited}

In \autoref{sect:ta}, we already outlined the timing argument (\autoref{eqn:rtvrvt}).  There are four equations and four unknowns ($a, e, M$ and $\eta$).  Apart from the special case $e = 1$, these equations must be solved numerically to yield multiple solutions corresponding to increasing numbers of pericentric passages.  To proceed, we combine the equations to eliminate $a$, $e$ and $M$:
\begin{equation}
    t = \frac{r \left( \left| \vr \right| \eta \seta \edenom - \vr^2 \sin^2
        \eta \right)} {\vr \left( \edenom - \left| \vr \right| \ceta
        \right) ^2}.
    \label{eqn:teta}
\end{equation}
To solve \autoref{eqn:teta} for $\eta$, we must provide $t$, $r$, $\vr$ and $\vt$ at the present time.  As before, we assume $t = \tu$, the age of the universe (see \autoref{sect:ta}).  The remainder of the parameters may be estimated from position and velocity data of M31 and the satellites.  The separations $r$ are calculated using trigonometry from the heliocentric distances and positions (\autoref{eqn:rsep}). The velocities $\vr$ and $\vt$ have contributions from both the line-of-sight velocities and the proper motions, both of which much be corrected for the solar peculiar motion, motion of the LSR and the relative motion of the Milky Way and M31 (see \autoref{sect:satprobs}).  However, as we do not have proper motion estimates for the satellites, we cannot compute either $\vr$ or $\vt$; instead we model to account for this incomplete data.

The corrected line-of-sight velocities $\vcor$ of the satellites in the M31 frame have contributions from both the radial velocities $\vr$ and the tangential velocities $\vt$ referred to the centre of M31,
\begin{equation}
    \vcor = \vr \cos \alpha + \vt \sin \alpha \cos \xi,
    \label{eqn:vcor}
\end{equation}
where $\alpha$ is the angle between the line joining the centre of M31 to the centre of the dSph and the line-of-sight to the dSph, and $\xi$ the angle between the tangential velocity vector and the plane containing M31, the dSph and the Sun. Now, $\alpha$ may be calculated via
\begin{equation}
    \cos \alpha = \frac{d^2 + r^2 - \dand^2}{2 d r}, \qquad \sin
        \alpha = \frac{\dand \sin \theta}{r},
\end{equation}
where $\theta$ is the angular separation of the dSph and M31 on the sky. It is convenient to introduce the (positive) velocity ratio $R = \vt / |\vr|$, such that
\begin{equation}
    \vt = \frac{R \vcor}{\sign \left( \vr \right) \cos \alpha
        + R \sin \alpha \cos \xi}.
\end{equation}
There is no available data on the velocity ratio $R$ and the angle of inclination of the orbit with respect to the plane of the sky $\xi$. Even so, the solution corresponding to the minimum mass $M$ is well-defined. A little consideration shows that on physical grounds, the orbit's kinetic energy -- and hence the total mass -- is minimised if $\xi =0$ (for $\vcor >0$) and if $\xi = \pi$ (for $\vcor <0$, which is the case for And\,XII and And\,XIV). A good approximation to the value of $R$ for the minimum mass is given by minimising the Lagrangian $L$
\begin{equation}
    L = \frac{1}{2} \left( \vr^2 + \vt^2 \right) - \ell \left( \vr \cos \alpha
        - \vt \sin \alpha - \vcor \right)
\end{equation}
where $\ell$ is the Lagrange multiplier. This leads to the solution $R = \tan \alpha$. In fact, though this is a good physical guide to the location of the solution, it is not exact. However, it provides a good starting point for a grid search, in which for successive $R$ values, \autoref{eqn:teta} is solved for $\eta$ iteratively and the minimum mass is located.


\subsection{Results}

\begin{table*}
\begin{center}
    \caption{Orbital parameters resulting from the timing argument analyses.}
    \label{table:results}
    \begin{tabular}{lccccccc}
        \hline
        \hline
        & solution & semi-major axis & eccentricity & mass
            & apocentre & pericentre & period \\
        & & (kpc) & & ($\times 10^{12} \Msun$) & (kpc) & (kpc) & (Gyr) \\
        \hline
        And\,XII & 1 & 364 & 0.90 & 2.1 & 690 & 38 & 14.3 \\
                 & 2 & 244 & 0.87 & 2.5 & 455 & 32 &  7.1 \\
                 & 3 & 196 & 0.86 & 2.9 & 364 & 28 &  4.7 \\
                 & 4 & 170 & 0.86 & 3.4 & 315 & 24 &  3.5 \\
        \hline
        And\,XIV & 1 & 292 & 0.49 & 1.1 & 436 & 148 & 14.2 \\
                 & 2 & 199 & 0.30 & 1.3 & 258 & 140 &  7.2 \\
                 & 3 & 163 & 0.20 & 1.6 & 195 & 130 &  4.9 \\
                 & 4 & 142 & 0.23 & 2.1 & 175 & 109 &  3.7 \\
        \hline
        \hline
    \end{tabular}
\end{center}
\end{table*}

We start by searching for a solution $\eta \in [\pi, 2 \pi]$, thus assuming that the dwarfs are on their first infall into M31 (multiple pericentric passages necessarily imply higher masses). For And\,XII we are able to place a lower constraint on the mass of the system of $M \ge 2.1 \times 10^{12} \Msun$, corresponding to $R = 0.6$.  Similarly for And\,XIV, there is a minimum mass estimate of $M \ge 1.1 \times 10^{12} \Msun$ when $R = 4.3$.  The corresponding orbital parameters are given in \autoref{table:results}.  Although our primary aim here is to understand the orbits of And\,XII and And\,XIV, we note that these are in reasonable agreement with earlier mass estimates $M = 1.23^{+1.8}_{-0.6} \times 10^{12} \Msun$ \citep{evans2000a,evans2000b} and $M = 1.5 \pm 0.4 \times 10^{12} \Msun$ \citep{watkins2010}. Remember that these are minimum mass solutions that we have adopted here; higher mass solutions exist, however moving away from this minimum would quickly result in unfeasibly high mass estimates for M31, so we are confident that the orbital parameters will be close to the values predicted by this analysis.  The resulting orbit of And\,XII is highly eccentric, it has an eccentricity $e \sim 0.9$ with a corresponding pericentre distance of only a few kiloparsecs.  The orbit of And\,XIV is of a more moderate eccentricity and puts And\,XIV very near pericentre at the current time, which would explain its high velocity.

\begin{figure}
\begin{center}
    \includegraphics[width=\linewidth]{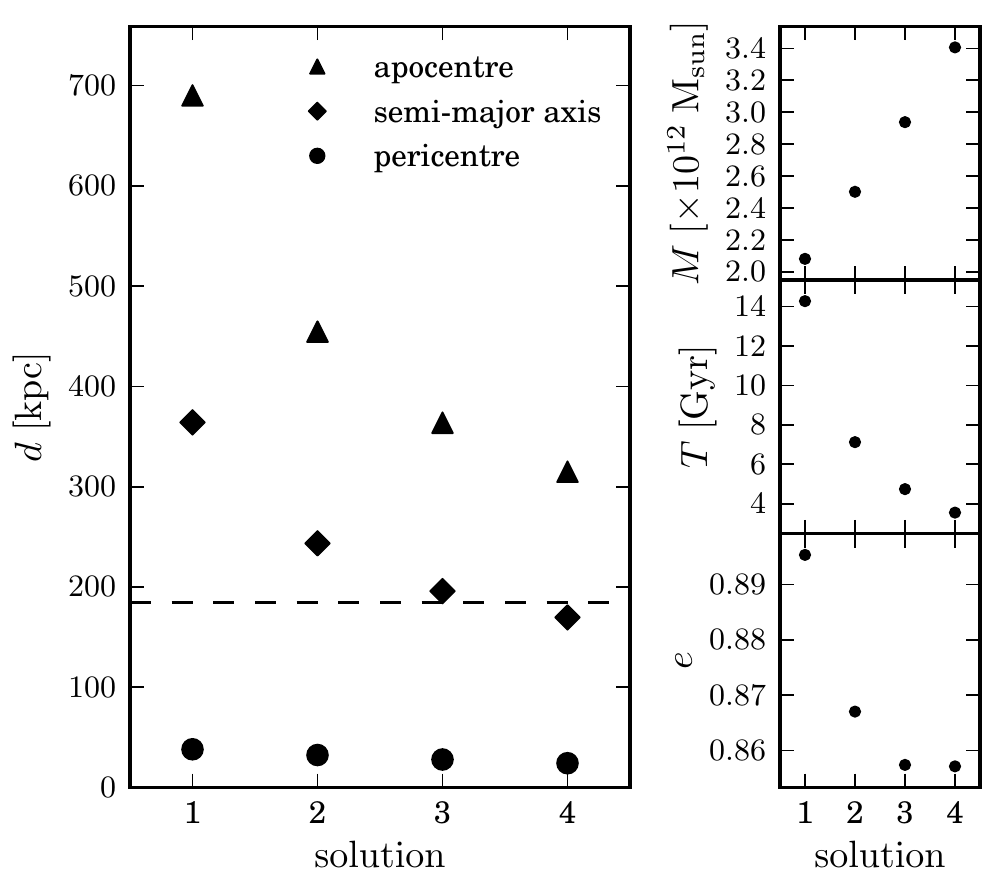}
    \includegraphics[width=\linewidth]{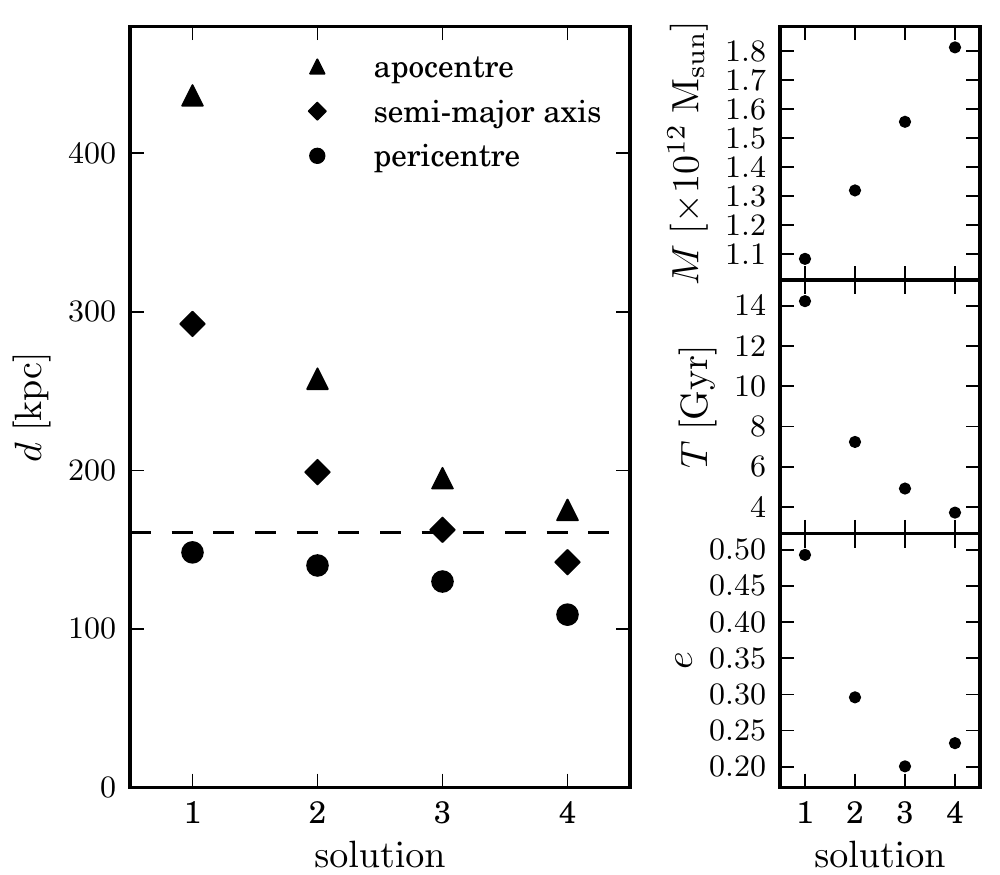}
    \caption{The first 4 $\eta$ solutions for And\,XII (top set) and And\,XIV (bottom set).  Left: apocentre (triangles), semi-major axis (diamonds) and pericentre (circles) as a function of solution number.  The dotted line shows the current position of the satellite.  It appears that the apocentre asymptotes to the current position of the satellite while the pericentre asymptotes to 0.  Top right: Minimum mass estimate $M$ as a function of solution number.  Middle right: orbital period $T$ as a function of solution number.  Bottom right: eccentricity $e$ as a function of solution number.}
    \label{fig:ta_solutions}
\end{center}
\end{figure}

We now repeat the analysis for solutions corresponding to multiple pericentric passages. The results for the first four solutions are shown in \autoref{fig:ta_solutions}, with the upper set of graphs pertaining to And\,XII and the lower set to And\,XIV.  The left panels shows the apocentres, pericentres and semi-major-axis lengths of the orbits. The top-right panels show that the mass estimate increases with increasing numbers of pericentric passages.  The corresponding orbital properties are given in \autoref{table:results}.  Again for And\,XII, all of these orbits are highly eccentric with small pericentre distances; this poses a problem for solutions that allow for more than one pericentre passage.  We would expect such close encounters with M31 to result in a significant amount of tidal distortion, but there is little evidence for such an interaction around either satellite.  Furthermore, the mass estimates for the system (already high for the first solution) become very high for subsequent solutions.  From this, we conclude that And\,XII cannot have passed through pericentre and, thus, must be on its first infall. For And\,XIV, it is less certain that we can rule out multiple-orbit scenarios; all four solutions predict masses consistent with previous estimates and the orbits are all reasonable on observational grounds.

\begin{figure}
\begin{center}
    \includegraphics[width=\linewidth]{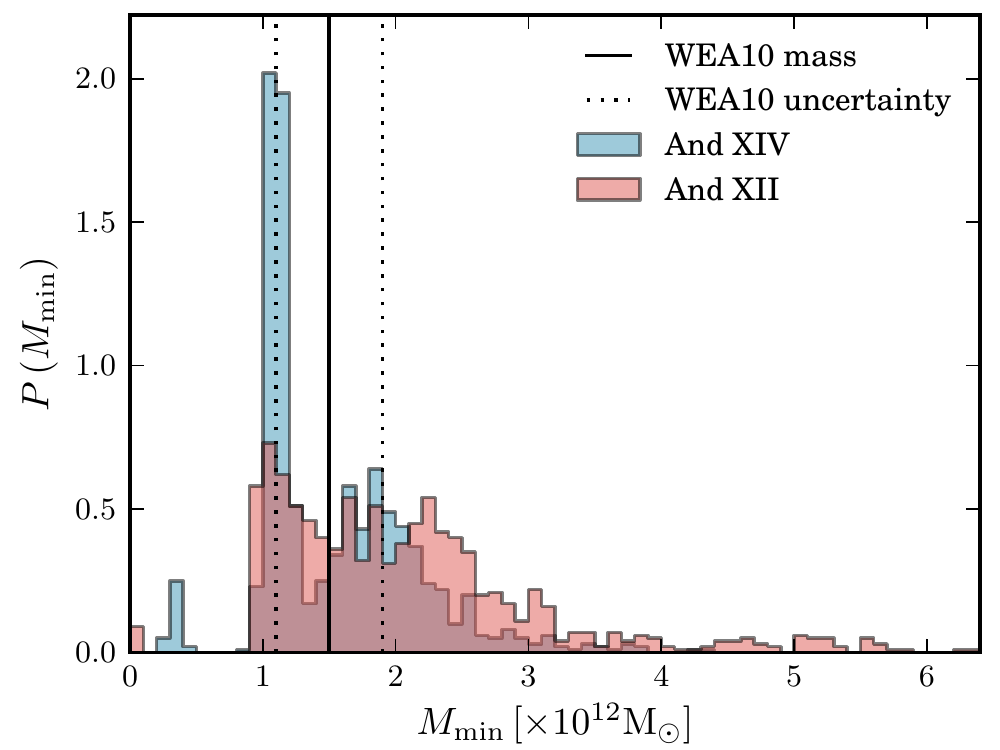}
    \caption{The distribution of minimum mass estimates for And\,XII (red histogram) and And\,XIV (blue histogram).  The solid and dotted lines show the best M31 mass estimate and the uncertainties from \citet{watkins2010}.}
    \label{fig:massdbns}
\end{center}
\end{figure}

So far, this analysis has used a single estimate for the separation distance and line-of-sight velocity of the satellite, and does not account for any uncertainties in these measurements, which we know to be considerable in some cases.  As discussed in \autoref{sect:data}, we now use full separation profiles and assume that the line-of-sight velocities have Gaussian errors.  We select $10^3$ separations $r$ and line-of-sight velocities $\vlos$ from these distributions and perform the same minimum-mass analysis, now requiring that the satellites are on their first infall.  The median and 1-sigma uncertainties in the minimum mass estimates are $M = 1.9_{-0.8}^{+1.0} \times 10^{12} \Msun$ for And\,XII and $M = 1.3_{-0.2}^{+0.8} \times 10^{12} \Msun$ for And\,XIV.  The full distributions of minimum mass estimates are shown in \autoref{fig:massdbns} with And\,XII in red and And\,XIV in blue; the solid (dotted) lines show the best estimate (uncertainties) from \citet{watkins2010}.  Both mass distributions are consistent with this mass estimate.  Indeed, the And\,XII distribution agrees very well with previous estimates, supporting the idea that And\,XII must be on its first infall into the M31 system.  The And\,XIV mass distribution, on the other hand, shows two distinct peaks. This is a direct consequence of the double-peaked heliocentric distance profile from \citet{conn2012}, such that smaller heliocentric distances produced larger host-satellite separations and thus, higher mass estimates. (Subsequent solutions assuming multiple pericentric passages produce minimum-mass profiles of similar shapes shifted to higher masses.) The low-mass peak contains low and moderate eccentricity orbits, with large pericentre distances, so solutions admitting multiple pericentre passages would be feasible; the high-mass peaks correspond to orbits of higher eccentricity and lower pericentre distance, such that we could make the same arguments as we did earlier for And\,XII and say that And\,XIV cannot have passed through pericentre.  As such, even in the absence of proper motion data, improved distance estimates are required to make any conclusion about And\,XIV via this analysis.


\section{Conclusions}
\label{sect:conclusions}

We have used the timing argument to study the orbits of the satellites of M31.  In all previous timing argument studies, the Milky Way has formed one half of the pair under consideration; never before has the timing argument been applied to two external galaxies as we do here. Two satellites, And\,XII and And\,XIV, are of particular interest as they are believed to be on their first infall into the M31 system as they have very high line-of-sight velocities, higher than would be expected at their respective distances from M31.

As we lack transverse velocity information for the satellites, we cannot recover their orbits exactly.  Instead we use Monte Carlo simulations to model the information that we do not have.  We generate a series of orbits from physically-motivated distribution function and ``observe" each orbit from a random viewing position.  We then use Bayesian analysis to determine the likelihood of each satellite given the set of orbits generated for each distribution function.

In order to apply the timing argument, we have assumed that the orbits are nearly-Keplerian, that the Milky Way and other satellites have a negligible effect on the orbits and that the orbits are part of a smooth, well-mixed population.  For the majority of the satellites, the models can recover the satellites very well and these assumptions hold.  However, And\,XII and And\,XIV are both very poorly fit by any of the models we have considered here; they are indeed unusual, as their velocities suggest.  \citet{li2008a} showed that the timing argument performs well on a host-satellite system, even with the presence of a nearby Milky-Way-like companion for the host, and their distances put them well within the bulk of the satellite population where we would not expect the nearly Keplerian assumption to break down.  Thus, we conclude that it is the final assumption that is incorrect, indicating that they are not well fit by a smooth distribution function.  From this it is reasonable to conclude that they are on their first infall into the M31 system, or that they are recent ejections from the system after a three-body encounter \citep{sales2007}.

In the light of this result, we relax the assumption that the orbits of And\,XII and And\,XIV are part of a well-mixed population and use the observed satellite properties to estimate their orbital parameters.  Again, due to the lack of transverse velocities, we cannot do this exactly; instead we are forced to search over a grid of possible transverse velocity measurements and adopt the minimum mass solution.  From this, we find that the And\,XII orbit solution is
realistic only when we assume that it is on its first infall; the orbit we predict is highly eccentric with a pericentre distance that will bring it close to the centre of M31.  For And\,XIV, we cannot rule out multiple pericentric passages; improved distance determination is needed to determine the most likely solution.  Both satellites predict M31 masses consistent with previous studies.

Finally, we consider whether the satellites might be infalling in groups.  We find four candidate groups (three pairs and one triple system), but further work is needed to confirm these predictions.  Our results favour an association between Cass\,II and NGC\,185, though not for NGC\,185 and NGC\,147, which have often been associated in the past.  Other possible candidates for group infall include the pair And\,I and And\,XVII; the pair And\,IX and And\,X; and the triple And\,V, And\,XXV and NGC\,147.

In this paper we have studied the M31 satellites. However, the method we have used is also readily applicable to the Milky Way satellite system.  It would be particularly interesting to compare the phase-space distribution function favoured by the Milky Way system with that predicted here, and to study the orbits of the Milky Way satellites with their M31 counterparts.  We could also use this method to investigate whether the ultrafaint dwarf spheroidals are kinematically different to the classical dwarf spheroidals.

We need not limit analyses of this nature to such massive hosts; we have already discussed the growing body of evidence that And\,XXII is not a satellite of M31 but of M33 and there are two extended clusters, EC\,1 and EC\,2, that are also found in close proximity to M33.  The techniques we have developed here could also be applied to this system to study their orbital properties and determine if the proposed satellites are truly bound to M33.

\qquad \\

\noindent We are grateful to Anthony Conn and Michelle Collins for making their distance profiles and velocity measurements available in advance of publication, and to David Hogg for interesting (and very useful) discussions.


\bibliographystyle{mn2e}
\bibliography{refs}

\begin{thebibliography}{}
\small
\itemindent -0.48cm

\bibitem[\protect\citeauthoryear{{Bell}, {Slater} \& {Martin}}{{Bell}  et~al.}{2011}]{bell2011}{Bell} E.~F.,  {Slater} C.~T.,    {Martin} N.~F.,  2011, \apjl, 742, L15

\bibitem[\protect\citeauthoryear{{Bender}, {Paquet} \& {Nieto}}{{Bender}  et~al.}{1991}]{bender1991}{Bender} R.,  {Paquet} A.,    {Nieto} J.-L.,  1991, \aap, 246, 349

\bibitem[\protect\citeauthoryear{{Brasseur}, {Martin}, {Rix}, {Irwin}, {Ferguson}, {McConnachie} \& {de Jong}}{Brasseur et~al.}{2011}]{brasseur2011}{Brasseur} C.~M.,  {Martin} N.~F.,  {Rix} H.-W.,  {Irwin} M.,  {Ferguson}  A.~M.~N.,  {McConnachie} A.~W.,    {de Jong} J.,  2011, \apj, 729, 23

\bibitem[\protect\citeauthoryear{{Chapman}, {Pe{\~n}arrubia}, {Ibata}, {McConnachie}, {Martin}, {Irwin}, {Blain}, {Lewis}, {Letarte}, {Lo}, {Ludlow} \& {O'neil}}{Chapman et~al.}{2007}]{chapman2007}{Chapman} S.~C.~{et al.},  2007, \apjl, 662, L79

\bibitem[\protect\citeauthoryear{{Chapman}, {Widrow}, {Collins}, {Dubinski}, {Ibata}, {Rich}, {Ferguson}, {Irwin}, {Lewis}, {Martin}, {McConnachie}, {Penarrubia} \& {Tanvir}}{Chapman et~al.}{2012}]{chapman2012}{Chapman} S.~C.~{et al.},  2012,  ArXiv e-prints

\bibitem[\protect\citeauthoryear{{Cole}, {Tolstoy}, {Gallagher}, {Hoessel}, {Mould}, {Holtzman}, {Saha}, {Ballester}, {Burrows}, {Clarke}, {Crisp}, {Griffiths}, {Grillmair}, {Hester}, {Krist}, {Meadows}, {Scowen}, {Stapelfeldt}, {Trauger}, {Watson} \& {Westphal}}{Cole et~al.}{1999}]{cole1999}{Cole} A.~A.~{et al.},  1999, \aj, 118, 1657

\bibitem[\protect\citeauthoryear{{Collins}, {Chapman}, {Irwin}, {Martin}, {Ibata}, {Zucker}, {Blain}, {Ferguson}, {Lewis}, {McConnachie} \& {Pe{\~n}arrubia}}{Collins et~al.}{2010}]{collins2010}{Collins} M.~L.~M.~{et al.},  2010, \mnras, 407, 2411

\bibitem[\protect\citeauthoryear{{Collins et al.}}{{Collins et  al.}}{2012}]{collins2012}{Collins et al.} 2012, submitted

\bibitem[\protect\citeauthoryear{{Conn}, {Ibata}, {Lewis}, {Parker}, {Zucker}, {Martin}, {McConnachie}, {Irwin}, {Tanvir}, {Fardal}, {Ferguson}, {Chapman} \& {Valls-Gabaud}}{Conn et~al.}{2012}]{conn2012}{Conn} A.~R.~{et al.},   2012, ArXiv e-prints

\bibitem[\protect\citeauthoryear{{Courteau} \& {van den Bergh}}{{Courteau} \&  {van den Bergh}}{1999}]{courteau1999}{Courteau} S.,  {van den Bergh} S.,  1999, \aj, 118, 337

\bibitem[\protect\citeauthoryear{{de Jong}, {Martin}, {Rix}, {Smith}, {Jin} \& {Macci{\`o}}}{de Jong et~al.}{2010}]{dejong2010}{de Jong} J.~T.~A.,  {Martin} N.~F.,  {Rix} H.,  {Smith} K.~W.,  {Jin} S.,  {Macci{\`o}} A.~V.,  2010, \apj, 710, 1664

\bibitem[\protect\citeauthoryear{{Dehnen} \& {Binney}}{{Dehnen} \&  {Binney}}{1998}]{dehnen1998}{Dehnen} W.,  {Binney} J.~J.,  1998, \mnras, 298, 387

\bibitem[\protect\citeauthoryear{{Evans}, {Hafner} \& {de Zeeuw}}{{Evans}  et~al.}{1997}]{evans1997}{Evans} N.~W.,  {Hafner} R.~M.,    {de Zeeuw} P.~T.,  1997, \mnras, 286, 315

\bibitem[\protect\citeauthoryear{{Evans} \& {Wilkinson}}{{Evans} \&  {Wilkinson}}{2000}]{evans2000a}{Evans} N.~W.,  {Wilkinson} M.~I.,  2000, \mnras, 316, 929

\bibitem[\protect\citeauthoryear{{Evans}, {Wilkinson}, {Guhathakurta}, {Grebel} \& {Vogt}}{Evans et~al.}{2000}]{evans2000b}{Evans} N.~W.,  {Wilkinson} M.~I.,  {Guhathakurta} P.,  {Grebel} E.~K.,  {Vogt} S.~S.,  2000, \apjl, 540, L9

\bibitem[\protect\citeauthoryear{{Fiorentino}, {Contreras Ramos}, {Tolstoy}, {Clementini} \& {Saha}}{Fiorentino et~al.}{2012}]{fiorentino2012}{Fiorentino} G.,  {Contreras Ramos} R.,  {Tolstoy} E.,  {Clementini} G.,  {Saha} A.,  2012, \aap, 539, A138

\bibitem[\protect\citeauthoryear{{Huchra}, {Vogeley} \& {Geller}}{{Huchra}  et~al.}{1999}]{huchra1999}{Huchra} J.~P.,  {Vogeley} M.~S.,    {Geller} M.~J.,  1999, \apjs, 121, 287

\bibitem[\protect\citeauthoryear{{Huchtmeier}, {Karachentsev} \& {Karachentseva}}{Huchtmeier et~al.}{2003}]{huchtmeier2003}{Huchtmeier} W.~K.,  {Karachentsev} I.~D.,    {Karachentseva} V.~E.,  2003,  \aap, 401, 483

\bibitem[\protect\citeauthoryear{{Ibata}, {Martin}, {Irwin}, {Chapman}, {Ferguson}, {Lewis} \& {McConnachie}}{Ibata et~al.}{2007}]{ibata2007}{Ibata} R.,  {Martin} N.~F.,  {Irwin} M.,  {Chapman} S.,  {Ferguson} A.~M.~N.,  {Lewis} G.~F.,    {McConnachie} A.~W.,  2007, \apj, 671, 1591

\bibitem[\protect\citeauthoryear{{Irwin}, {Ferguson}, {Huxor}, {Tanvir}, {Ibata} \& {Lewis}}{Irwin et~al.}{2008}]{irwin2008}{Irwin} M.~J.,  {Ferguson} A.~M.~N.,  {Huxor} A.~P.,  {Tanvir} N.~R.,  {Ibata}  R.~A.,    {Lewis} G.~F.,  2008, \apjl, 676, L17

\bibitem[\protect\citeauthoryear{{Irwin et al.}}{{Irwin et  al.}}{2012}]{irwin2012}{Irwin et al.} 2012, in prep

\bibitem[\protect\citeauthoryear{{Jensen}, {Tonry}, {Barris}, {Thompson}, {Liu}, {Rieke}, {Ajhar} \& {Blakeslee}}{Jensen et~al.}{2003}]{jensen2003}{Jensen} J.~B.,  {Tonry} J.~L.,  {Barris} B.~J.,  {Thompson} R.~I.,  {Liu}  M.~C.,  {Rieke} M.~J.,  {Ajhar} E.~A.,    {Blakeslee} J.~P.,  2003, \apj,  583, 712

\bibitem[\protect\citeauthoryear{{Kahn} \& {Woltjer}}{{Kahn} \&  {Woltjer}}{1959}]{kahn1959}{Kahn} F.~D.,  {Woltjer} L.,  1959, \apj, 130, 705

\bibitem[\protect\citeauthoryear{{Kalirai}, {Beaton}, {Geha}, {Gilbert}, {Guhathakurta}, {Kirby}, {Majewski}, {Ostheimer}, {Patterson} \& {Wolf}}{Kalirai et~al.}{2010}]{kalirai2010}{Kalirai} J.~S.~{et al.},  2010, \apj, 711, 671

\bibitem[\protect\citeauthoryear{{Kallivayalil}, {Besla}, {Sanderson} \& {Alcock}}{Kallivayalil et~al.}{2009}]{kallivayalil2009}{Kallivayalil} N.,  {Besla} G.,  {Sanderson} R.,    {Alcock} C.,  2009, \apj,  700, 924

\bibitem[\protect\citeauthoryear{{Karachentsev}, {Karachentseva}, {Huchtmeier} \& {Makarov}}{Karachentsev et~al.}{2004}]{karachentsev2004}{Karachentsev} I.~D.,  {Karachentseva} V.~E.,  {Huchtmeier} W.~K.,    {Makarov}  D.~I.,  2004, \aj, 127, 2031

\bibitem[\protect\citeauthoryear{{Kerr} \& {Lynden-Bell}}{{Kerr} \&  {Lynden-Bell}}{1986}]{kerr1986}{Kerr} F.~J.,  {Lynden-Bell} D.,  1986, \mnras, 221, 1023

\bibitem[\protect\citeauthoryear{{Li} \& {Helmi}}{{Li} \&  {Helmi}}{2008}]{li2008b}{Li} Y.,  {Helmi} A.,  2008, \mnras, 385, 1365

\bibitem[\protect\citeauthoryear{{Li} \& {White}}{{Li} \&  {White}}{2008}]{li2008a}{Li} Y.,  {White} S.~D.~M.,  2008, \mnras, 384, 1459

\bibitem[\protect\citeauthoryear{{Majewski}, {Beaton}, {Patterson}, {Kalirai}, {Geha}, {Mu{\~n}oz}, {Seigar}, {Guhathakurta}, {Gilbert}, {Rich}, {Bullock} \& {Reitzel}}{Majewski et~al.}{2007}]{majewski2007}{Majewski} S.~R.~{et al.},  2007,  \apjl, 670, L9

\bibitem[\protect\citeauthoryear{{Martin}, {Ibata}, {Irwin}, {Chapman}, {Lewis}, {Ferguson}, {Tanvir} \& {McConnachie}}{Martin et~al.}{2006}]{martin2006}{Martin} N.~F.,  {Ibata} R.~A.,  {Irwin} M.~J.,  {Chapman} S.,  {Lewis} G.~F.,  {Ferguson} A.~M.~N.,  {Tanvir} N.,    {McConnachie} A.~W.,  2006, \mnras,  371, 1983

\bibitem[\protect\citeauthoryear{{Martin}, {McConnachie}, {Irwin}, {Widrow}, {Ferguson}, {Ibata}, {Dubinski}, {Babul}, {Chapman}, {Fardal}, {Lewis}, {Navarro} \& {Rich}}{Martin et~al.}{2009}]{martin2009}{Martin} N.~F.~{et al.},  2009, \apj,  705, 758

\bibitem[\protect\citeauthoryear{{McConnachie}, {Huxor}, {Martin}, {Irwin}, {Chapman}, {Fahlman}, {Ferguson}, {Ibata}, {Lewis}, {Richer} \& {Tanvir}}{McConnachie et~al.}{2008}]{mcconnachie2008}{McConnachie} A.~W.~{et al.},  2008, \apj, 688, 1009

\bibitem[\protect\citeauthoryear{{McConnachie}, {Irwin}, {Ferguson}, {Ibata}, {Lewis} \& {Tanvir}}{McConnachie et~al.}{2005}]{mcconnachie2005}{McConnachie} A.~W.,  {Irwin} M.~J.,  {Ferguson} A.~M.~N.,  {Ibata} R.~A.,  {Lewis} G.~F.,    {Tanvir} N.,  2005, \mnras, 356, 979

\bibitem[\protect\citeauthoryear{{Monachesi}, {Trager}, {Lauer}, {Freedman}, {Dressler}, {Grillmair} \& {Mighell}}{Monachesi et~al.}{2011}]{monachesi2011}{Monachesi} A.,  {Trager} S.~C.,  {Lauer} T.~R.,  {Freedman} W.,  {Dressler}  A.,  {Grillmair} C.,    {Mighell} K.~J.,  2011, \apj, 727, 55

\bibitem[\protect\citeauthoryear{{Richardson}, {Irwin}, {McConnachie}, {Martin}, {Dotter}, {Ferguson}, {Ibata}, {Chapman}, {Lewis}, {Tanvir} \& {Rich}}{Richardson et~al.}{2011}]{richardson2011}{Richardson} J.~C.~{et al.},  2011, \apj, 732, 76

\bibitem[\protect\citeauthoryear{{Sakai}, {Madore} \& {Freedman}}{{Sakai}  et~al.}{1999}]{sakai1999}{Sakai} S.,  {Madore} B.~F.,    {Freedman} W.~L.,  1999, \apj, 511, 671

\bibitem[\protect\citeauthoryear{{Sales}, {Navarro}, {Abadi} \& {Steinmetz}}{Sales et~al.}{2007}]{sales2007}{Sales} L.~V.,  {Navarro} J.~F.,  {Abadi} M.~G.,    {Steinmetz} M.,  2007,  \mnras, 379, 1475

\bibitem[\protect\citeauthoryear{{Sarajedini}, {Yang}, {Monachesi}, {Lauer} \& {Trager}}{Sarajedini et~al.}{2012}]{sarajedini2012}{Sarajedini} A.,  {Yang} S.-C.,  {Monachesi} A.,  {Lauer} T.~R.,    {Trager}  S.~C.,  2012, \mnras, 425, 1459

\bibitem[\protect\citeauthoryear{{Slater}, {Bell} \& {Martin}}{{Slater}  et~al.}{2011}]{slater2011}{Slater} C.~T.,  {Bell} E.~F.,    {Martin} N.~F.,  2011, \apjl, 742, L14

\bibitem[\protect\citeauthoryear{{Spergel}, {Bean}, {Dore}, {Nolta}, {Bennett}, {Dunkley}, {Hinshaw}, {Jarosik}, {Komatsu}, {Page}, {Peiris}, {Verde}, {Halpern}, {Hill}, {Kogut}, {Limon}, {Meyer}, {Odegard}, {Tucker}, {Weiland}, {Wollack} \& {Wright}}{Spergel et~al.}{2007}]{spergel2007}{Spergel} D.~N.~{et al.},  2007, \apjs, 170, 377

\bibitem[\protect\citeauthoryear{{Strigari}, {Bullock}, {Kaplinghat}, {Simon}, {Geha}, {Willman} \& {Walker}}{Strigari et~al.}{2008}]{strigari2008}{Strigari} L.~E.,  {Bullock} J.~S.,  {Kaplinghat} M.,  {Simon} J.~D.,  {Geha}  M.,  {Willman} B.,    {Walker} M.~G.,  2008, \nat, 454, 1096

\bibitem[\protect\citeauthoryear{{Tollerud}, {Beaton}, {Geha}, {Bullock}, {Guhathakurta}, {Kalirai}, {Majewski}, {Kirby}, {Gilbert}, {Yniguez}, {Patterson}, {Ostheimer}, {Cooke}, {Dorman}, {Choudhury} \& {Cooper}}{Tollerud et~al.}{2012}]{tollerud2012}{Tollerud} E.~J.~{et al.},  2012, \apj,  752, 45

\bibitem[\protect\citeauthoryear{{Tonry}, {Dressler}, {Blakeslee}, {Ajhar}, {Fletcher}, {Luppino}, {Metzger} \& {Moore}}{Tonry et~al.}{2001}]{tonry2001}{Tonry} J.~L.,  {Dressler} A.,  {Blakeslee} J.~P.,  {Ajhar} E.~A.,  {Fletcher}  A.~B.,  {Luppino} G.~A.,  {Metzger} M.~R.,    {Moore} C.~B.,  2001, \apj,  546, 681

\bibitem[\protect\citeauthoryear{{van der Marel}, {Fardal}, {Besla}, {Beaton}, {Sohn}, {Anderson}, {Brown} \& {Guhathakurta}}{van der Marel et~al.}{2012}]{vandermarel2012}{van der Marel} R.~P.,  {Fardal} M.,  {Besla} G.,  {Beaton} R.~L.,  {Sohn}  S.~T.,  {Anderson} J.,  {Brown} T.,    {Guhathakurta} P.,  2012, \apj, 753, 8

\bibitem[\protect\citeauthoryear{{van der Marel} \& {Guhathakurta}}{{van der  Marel} \& {Guhathakurta}}{2008}]{vandermarel2008}{van der Marel} R.~P.,  {Guhathakurta} P.,  2008, \apj, 678, 187

\bibitem[\protect\citeauthoryear{{Watkins}, {Evans} \& {An}}{{Watkins}  et~al.}{2010}]{watkins2010}{Watkins} L.~L.,  {Evans} N.~W.,    {An} J.~H.,  2010, \mnras, 406, 264

\bibitem[\protect\citeauthoryear{{Zucker}, {Kniazev}, {Bell}, {Mart{\'{\i}}nez-Delgado}, {Grebel}, {Rix}, {Rockosi}, {Holtzman}, {Walterbos}, {Annis}, {York}, {Ivezi{\'c}}, {Brinkmann}, {Brewington}, {Harvanek}, {Hennessy}, {Kleinman}, {Krzesinski}, {Long}, {Newman}, {Nitta} \& {Snedden}}{Zucker et~al.}{2004}]{zucker2004}{Zucker} D.~B.~{et al.},  2004, \apjl, 612, L121

\bibitem[\protect\citeauthoryear{{Zucker}, {Kniazev}, {Mart{\'{\i}}nez-Delgado}, {Bell}, {Rix}, {Grebel}, {Holtzman}, {Walterbos}, {Rockosi}, {York}, {Barentine}, {Brewington}, {Brinkmann}, {Harvanek}, {Kleinman}, {Krzesinski}, {Long}, {Neilsen} Jr., {Nitta} \& {Snedden}}{Zucker et~al.}{2007}]{zucker2007}{Zucker} D.~B.~{et al.},  2007, \apjl, 659,  L21

\end{thebibliography}


\appendix

\section{Distance to M32}
\label{sect:distm32}

There have been a number of attempts to measure the heliocentric distance of M32 in recent years, using a variety of different methods, including RR Lyrae variables, surface brightness fluctuations and Red Clump stars.  These distance modulus estimates are given in \autoref{table:m32dm}, along with the method that was used to obtain the estimates and their references.  Faced with so many estimates, all of which result from careful, reliable studies, it is not clear which estimate should be adopted as the distance to M32; indeed different authors adopt different distance estimates in any M32 studies they undertake, and in some cases, studies simply adopt the same distance for M32 as they take for M31.

\begin{table}
\begin{center}
    \caption{M32 distance modulus estimates; the method used to obtain each estimate is also given, along with the source.}
    \label{table:m32dm}
    \begin{tabular}{ccc}
        \hline
        \hline
        $\mu \pm \sigma$ & method & reference\\
        \hline
        24.55 $\pm$ 0.08 & SBF & \citet{tonry2001} \\
        24.39 $\pm$ 0.08 & SBF & \citet{jensen2003} \\
        24.53 $\pm$ 0.12 & Red Clump & \citet{monachesi2011} \\
        24.42 $\pm$ 0.12 & RR Lyraes & \citet{sarajedini2012} \\
        24.33 $\pm$ 0.21 & RR Lyraes & \citet{fiorentino2012} \\
        \hline
        \hline
    \end{tabular}
\end{center}
\end{table}

All of these different methods have their advantages and disadvantages, with no one method being preferred over another; on the contrary, we are lucky to have so many different estimates from so many different data sets.  Taken together, the estimates complement and support each other, and their overall agreement is remarkable.  Instead of picking one distance estimate and ignoring the rest, we would like to use a value that takes into account all of the information in \autoref{table:m32dm}.  To that end, here we present a statistical analysis that attempts to combine the previous estimates to obtain an improved distance estimate.

The most straightforward option would be to take a simple mean and use the standard deviation as the uncertainty.  However, this does not take into account any uncertainties on the estimates themselves, only their scatter, and can be unreliable when the number of estimates is very small.  A weighted mean - using weights equal to the inverse square of the uncertainties - would neatly address the first point, but not the second.  Instead we turn to blind deconvolution, which is robust even with a small number of original estimates.  We can also fit for an extra, as-yet-ignored systematic error on all the measurements.

\begin{figure}
\begin{center}
    \includegraphics[width=\linewidth]{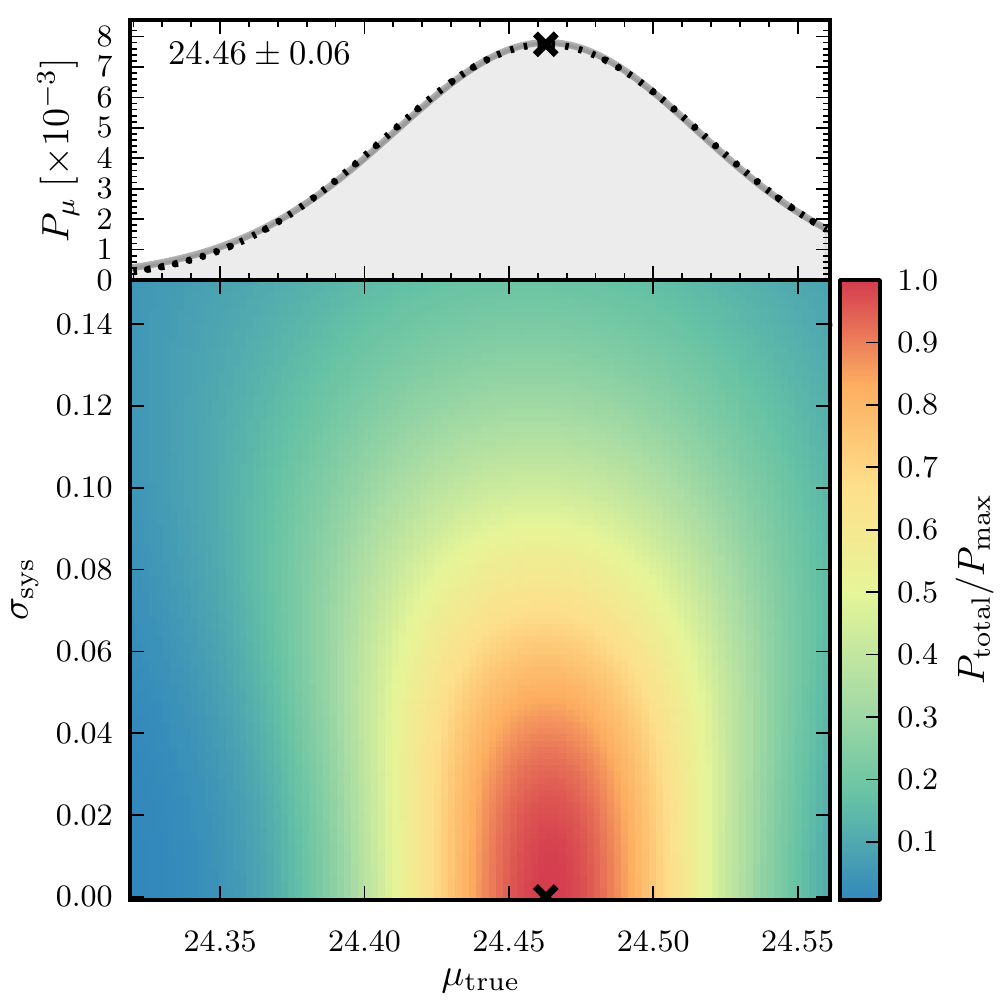}
    \caption{Bottom panel: The probability of observing the measured M32 distance moduli given their error estimates as a function of the true distance modulus $\mutrue$ and the systematic error $\sigsys$.  Top panel: The marginalised probability for true distance modulus $\mutrue$.  The point at which the probability is maximised is marked with a cross.}
    \label{fig:m32_grid}
\end{center}
\end{figure}

To start, consider a set of distance-modulus estimates $\mu_i$, each with uncertainty $\sigma_i$.  Let us assume that there is a true distance modulus $\mutrue$.  Let us further assume that there is a systematic error $\sigsys$ on each measurement that has not yet been accounted for in the quoted measurements such that the uncertainties on the measurements become
\begin{equation}
    \sigma_i^* = \sqrt{ \sigma_i^2 + \sigsys^2 }.
\end{equation}

Now the probability of observed measurement $\mu_i$ given the uncertainties and the true value is given by
\begin{equation}
    P \left( \mu_i | \sigma_i, \sigsys, \mutrue \right) = \frac{1}{\sqrt{2 \pi}
        \sigma_i^*} \exp \left( -\frac{\left( \mu_i - \mutrue \right)^2}
        {2 \sigma_i^{*2}} \right).
\end{equation}

The probability of obtaining all measurements $\{\mu_i\}$ is then
\begin{eqnarray}
    & & P \left( \left\{ \mu_i \right\}_{i=1}^N | \left\{ \sigma_i
        \right\}_{i=1}^N, \sigsys, \mutrue \right) \nonumber \\
    & & = \prod_{i=1}^N P \left(
        \mu_i | \sigma_i, \sigsys, \mutrue \right) \nonumber \\
    & & = \prod_{i=1}^N
        \frac{1}{\sqrt{2 \pi} \sigma_i^*} \exp \left( -\frac{\left( \mu_i -
        \mutrue \right)^2}{2 \sigma_i^{*2}} \right).
\end{eqnarray}

We calculate this probability over a grid of models with $\mutrue \in [24.32,24.56]$ and $\sigsys \in [0,0.15]$.  The results of this grid search are shown in the bottom panel of \autoref{fig:m32_grid}, the colours of the pixels indicate the probability of a given $(\mutrue,\sigsys)$ pair, with red representing high probability and blue representing low probability.  The distribution peaks at a value of $\sigsys$ close to 0, implying that there are no further systematic errors that must be taken into account in this case.  The statistical error $\sigsys$ is a nuisance parameter here; we must consider the possibility that there is an extra source of error on the estimates, however we are not interested in its value.  It is the value of $\mutrue$ that we wish to determine.  So we marginalise over $\sigsys$ (top panel of \autoref{fig:m32_grid}) and adopt the value of $\mutrue$ at which the marginalised probability is maximised as the best estimate for the distance modulus.  From the marginalised distribution for $\mutrue$, we obtain an estimate $\mutrue = 24.46 \pm 0.06$, which corresponds to a distance of $d = 781 \pm 20$~kpc.


\section{Separation-distance distributions}

\begin{figure*}
\begin{center}
    \includegraphics[width=\linewidth]{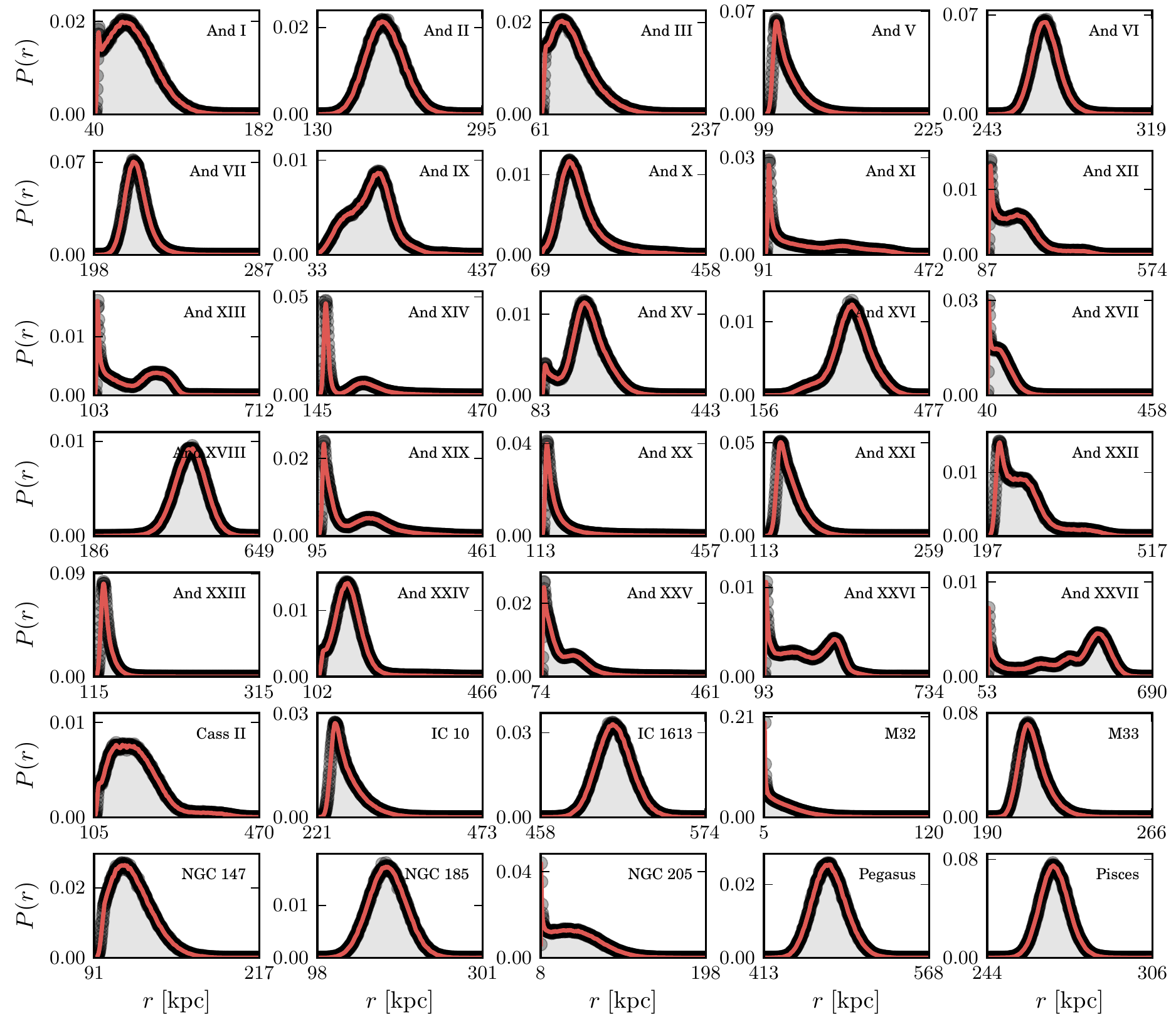}
    \caption{The distribution of separation distances for each of the M31 satellites.  For satellites that are far enough away from the centre of M31 or have small errors on their distance, the Gaussian errors on the heliocentric distance result in Gaussian errors on the separation distances.  For other satellite, the distribution finds a minimum separation distance and the distribution becomes distinctly non-Gaussian.  The red lines show the functions fitted to the distributions.}
    \label{fig:pr_full}
\end{center}
\end{figure*}

\autoref{fig:pr_full} shows the complete set of separation-distance distributions that were used for this analysis.

\label{lastpage}

\end{document}